\documentclass[aps,prb,reprint,amsmath,superscriptaddress]{revtex4-1}   
\usepackage{graphicx}
\usepackage{siunitx}
\usepackage{hyperref}

\begin{document}
	\title{Ferroelectric order versus metallicity in Sr$_{1-x}$Ca$_x$TiO$_{3-\delta}$ ($x=0.009$)}
	\date{\today}
	\author{Johannes Engelmayer}
	\author{Xiao Lin}
	\altaffiliation[Present address: ]{School of Science, Westlake University, 18 Shilongshan Road, 310024, Hangzhou, China}
	\author{Fulya Ko\c{c}}
	\author{Christoph P. Grams}
	\author{Joachim Hemberger}
	\affiliation{II. Physikalisches Institut, Universit\"{a}t zu K\"{o}ln, Z\"{u}lpicher Str. 77, 50937 K\"{o}ln, Germany}
	\author{Kamran Behnia}
	\affiliation{Laboratoire de Physique et d'\'{E}tude des Mat\'{e}riaux (UMR 8213 CNRS-ESPCI), PSL Research University, 10 Rue Vauquelin, 75005 Paris, France}
	\affiliation{II. Physikalisches Institut, Universit\"{a}t zu K\"{o}ln, Z\"{u}lpicher Str. 77, 50937 K\"{o}ln, Germany}
	\author{Thomas Lorenz}
	\email{tl@ph2.uni-koeln.de}
	\affiliation{II. Physikalisches Institut, Universit\"{a}t zu K\"{o}ln, Z\"{u}lpicher Str. 77, 50937 K\"{o}ln, Germany}
	
	\begin{abstract}
			We report on a thermal-expansion study of the ferroelectric phase transition in insulating Sr$_{1-x}$Ca$_x$TiO$_3$ ($x=0.009$) and its evolution upon increasing charge-carrier concentration up to $n\simeq\SI{60e19}{cm^{-3}}$. Although electric polarization is screened by mobile charge carriers, we find clear signatures of the ferroelectric phase transition in the thermal-expansion coefficient $\alpha$ of the weakly doped metallic samples. Upon increasing $n$, the transition temperature $T_\mathrm{C}(n)$ and the magnitude of the anomalies in $\alpha$ rapidly decrease up to a threshold carrier density $n^\star$ above which broadened anomalies remain present. There is no indication for a sign change of $\alpha$ as is expected for a pressure-dependent quantum phase transition with $n$ as the control parameter. Thus, the ferroelectriclike transition is either continuously fading away or it transforms to another low-temperature phase above $n^\star$, but this change hardly affects the temperature-dependent $\alpha(T)$ data. 
	\end{abstract}

	\maketitle

	\section{Introduction}
	
	Perovskite titanates $A$TiO$_3$ with divalent $A$-site ions contain tetravalent titanium with an empty $3d$ shell, such that these materials typically form band insulators. Some members of this $A^{2+}$Ti$^{4+}$O$_3$ family are ferroelectric such as BaTiO$_3$, PbTiO$_3$, CdTiO$_3$~\cite{VonHippel1950,Bhide1962,Sun1998,Kennedy2011}, whereas others like SrTiO$_3$, CaTiO$_3$, and EuTiO$_3$ show quantum paraelectric behavior~\cite{Nakamura1997,Kim1992,Lemanov1999,Mueller1979,Katsufuji2001,Kamba2007,Engelmayer2019}, i.e., a ferroelectric long-range order is suppressed by quantum fluctuations.
	Although a ferroelectric transition is absent in both SrTiO$_3$ and CaTiO$_3$~\cite{Hulm1950,Kim1992}, mixing Sr and Ca on the $A$ site (Sr$_{1-x}$Ca$_x$TiO$_3$) induces ferroelectricity already for tiny calcium substitutions $ 0.0018 \leq x< 0.02$ with increasing Curie temperature $T_\mathrm{C}(x)$. For larger $x$, relaxor ferroelectric behavior is observed and finally the material becomes antiferroelectric above $x\simeq  0.12$~\cite{Bednorz1984,Ranjan2000,Ranjan2001a,Ranjan2001b,Mishra2002}. 
	The high-temperature structure of Sr$_{1-x}$Ca$_x$TiO$_3$ is cubic (space group $Pm\bar{3}m$, No.~221), but upon cooling it changes to the tetragonal, centrosymmetric space group $I4/mcm$~(No.~140) at an $x$-dependent transition temperature $T_\mathrm{s}(x)$, which is significantly larger than the ferroelectric ordering temperature $T_\mathrm{C}$. Because the finite polarization in the ferroelectric state requires the absence of an inversion center, the transition at $T_\mathrm{C}$ necessarily involves a further symmetry reduction. The crystal structure in the ferroelectric phase was found to belong to the orthorhombic point group $mm2$~\cite{Bianchi1994,Kleemann1997,Carpenter2006} and for $x=0.02,0.04$ the space group $Ic2m$~(No.~46) was specified by x-ray diffraction~\cite{Mishra2009}.
	
	Charge-carrier doping in pure SrTiO$_3$ by, e.g., a partial removal of oxygen or substitution of Sr~(Ti) by La~(Nb), induces metallic conductivity~\cite{Frederikse1964,Tufte1967,Frederikse1967,Lee1975,Spinelli2010} and, for certain carrier concentrations, even superconductivity~\cite{Schooley1964,Lin2014}.
	In systems with both, calcium substitution and electron doping (Sr$_{1-x}$Ca$_x$TiO$_{3-\delta}$), the $T_\mathrm{C}$-related anomalies of the ferroelectric insulating parent compound persist within the metallic and superconducting phase~\cite{DeLima2015,Rischau2017}. Rischau~{\it et~al.}~investigated the evolution of this ferroelectriclike transition with charge-carrier concentration $n$ for Sr$_{1-x}$Ca$_x$TiO$_{3-\delta}$ with $x=0.0022$ and $x=0.009$~\cite{Rischau2017}.
	Based on minima in the resistivity data $\rho(T)$, a decreasing $T_\mathrm{C}$ upon increasing $n$ was derived and a disappearance of the ferroelectriclike phase above a critical carrier density $n_c$ that depends on the calcium content $x$. The mechanism behind this behavior remained unclear, but it was suggested that it could result from Friedel oscillations causing destructive interference of Ca-induced dipoles~\cite{Rischau2017}. In fact, such a mechanism was discussed theoretically already much earlier by Glinchuk~{\it et~al.}~\cite{Glinchuk1992,Glinchuk1994}. 
	
	Rowley~{\it et~al.}~discussed the appearance of ferroelectric order in insulating quantum paraelectrics in the context of quantum criticality~\cite{Rowley2014} where for SrTiO$_3$
	the quantum control parameter can be tuned either by stress~\cite{Uwe1976}, by chemical substitution like in Sr$_{1-x}$Ca$_x$TiO$_3$, or by oxygen-isotope exchange~\cite{Itoh1999}.
	In such a scenario, the charge-carrier concentration $n$ represents an additional control parameter towards a metallic ground state.
	The presence of a quantum phase transition is intrinsically tied to a diverging Gr\"{u}neisen parameter~\cite{Zhu2003}.
	For pressure-dependent quantum phase transitions, this holds for the Grüneisen ratio $\Gamma=\alpha/c_p$ with the thermal-expansion coefficient $\alpha$ and the specific heat $c_p$. In the vicinity of a quantum critical point $\Gamma$ exhibits a sign change~\cite{Garst2005}, which results from a sign change of $\alpha$, because $c_p$ is always positive~\footnote{The divergence and sign change of $\alpha/c_p$ require a finite pressure dependence of the underlying quantum phase transition, which, for example, may result from a pressure-dependent quantum critical magnetic field.}.
	Experimentally, such sign changes of $\alpha$ are observed in diverse materials where the control parameter is either a magnetic field~\cite{Kuechler2003,Gegenwart2006,Baier2007,Lorenz2007,Lorenz2008,Breunig2013,Breunig2017a}, 
	a chemical or hydrostatic pressure~\cite{Grube2018}, or the charge-carrier concentration~\cite{Meingast2012}.
	
	Here, we present a detailed study of the evolution of the ground state of Sr$_{1-x}$Ca$_x$TiO$_3$ with $x=0.009$ as a function of the charge-carrier density $n$ varying from the insulating parent compound to $n\simeq\SI{6E20}{cm^{-3}}$. Based on thermal-expansion measurements, we investigate the evolution of both the structural transition temperature $T_s$ and the $T_\mathrm{C}$-related transition as a function of carrier concentration $n$.
	With increasing $n$, $T_s$ essentially linearly decreases and the structural transition remains well defined over almost the entire doping range. In contrast, $T_\mathrm{C}(n)$ and the corresponding anomalies in $\alpha$ rapidly decrease in the low-$n$ range up to a threshold carrier concentration $n^\star$, above which broadened anomalies remain present up to the highest $n$. There is no evidence for a sign change in $\alpha$ as a function of $n$. This either suggests the absence of a sharp quantum phase transition, because the ferroelectriclike transition in Sr$_{1-x}$Ca$_x$TiO$_3$ is continuously vanishing, e.g.\ by varying from long range to short range, or the symmetry of the low-temperature order changes at some critical charge-carrier content $n^\star$ without being directly reflected in the macroscopic uniaxial expansion $\alpha(T)$. 
	
	The discussion of our results is split into two parts. Section \ref{pristine} discusses the symmetry changes at the structural and the ferroelectric transition of the pristine insulating material in order to clarify how the macroscopic uniaxial expansion $\alpha(T)$ is related to a (partial) twinning occurring at both transitions, which sets the basis for the discussion of the evolution of both transitions as a function of the charge-carrier concentration in \ref{doped}.

\section{Experimental}
	
	A commercial Sr$_{1-x}$Ca$_x$TiO$_3$ single crystal with $x=0.009$ was used for this study. The nominal Ca content was confirmed by secondary ion mass spectrometry (SIMS) as described in~\cite{DeLima2015}. The crystal was cut into cuboid pieces with all faces being cubic $\{100\}$ planes and dimensions optimized for Hall effect measurements (typically $0.5\times 2.5\times\SI{5}{mm}$).
	In order to induce electron doping the samples were annealed under vacuum ($\lesssim\SI{E-5}{mbar}$) for 1 to 2 hours at temperatures between \SI{700}{\celsius} and \SI{1000}{\celsius} depending on the intended charge-carrier concentration.
	Resistivity and Hall effect measurements were carried out via a standard six-probe method using the commercial physical property measurement system (\textsc{PPMS} by \textsc{Quantum Design}) and most of these data have been published in Ref.~\cite{Rischau2017}. On the same samples we studied the uniaxial thermal expansion by measuring the length change $\Delta L(T)$ using a home-built capacitance dilatometer. The samples were continuously heated from liquid-helium temperature to \SI{180}{K} at a rate of about \SI{0.1}{K\,min^{-1}} and the thermal-expansion coefficient $\alpha=(1/L_0)(\partial\Delta L/\partial T)$ was determined numerically. The heat-capacity measurement of pristine Sr$_{1-x}$Ca$_x$TiO$_3$ was performed using the microcalorimeter option of the PPMS.

	\section{Results and discussion}
	\label{result} 
	
	\subsection{Pristine Sr$_{0.991}$Ca$_{0.009}$TiO$_3$}
	\label{pristine}

\begin{figure}
		\includegraphics[width=1\columnwidth]{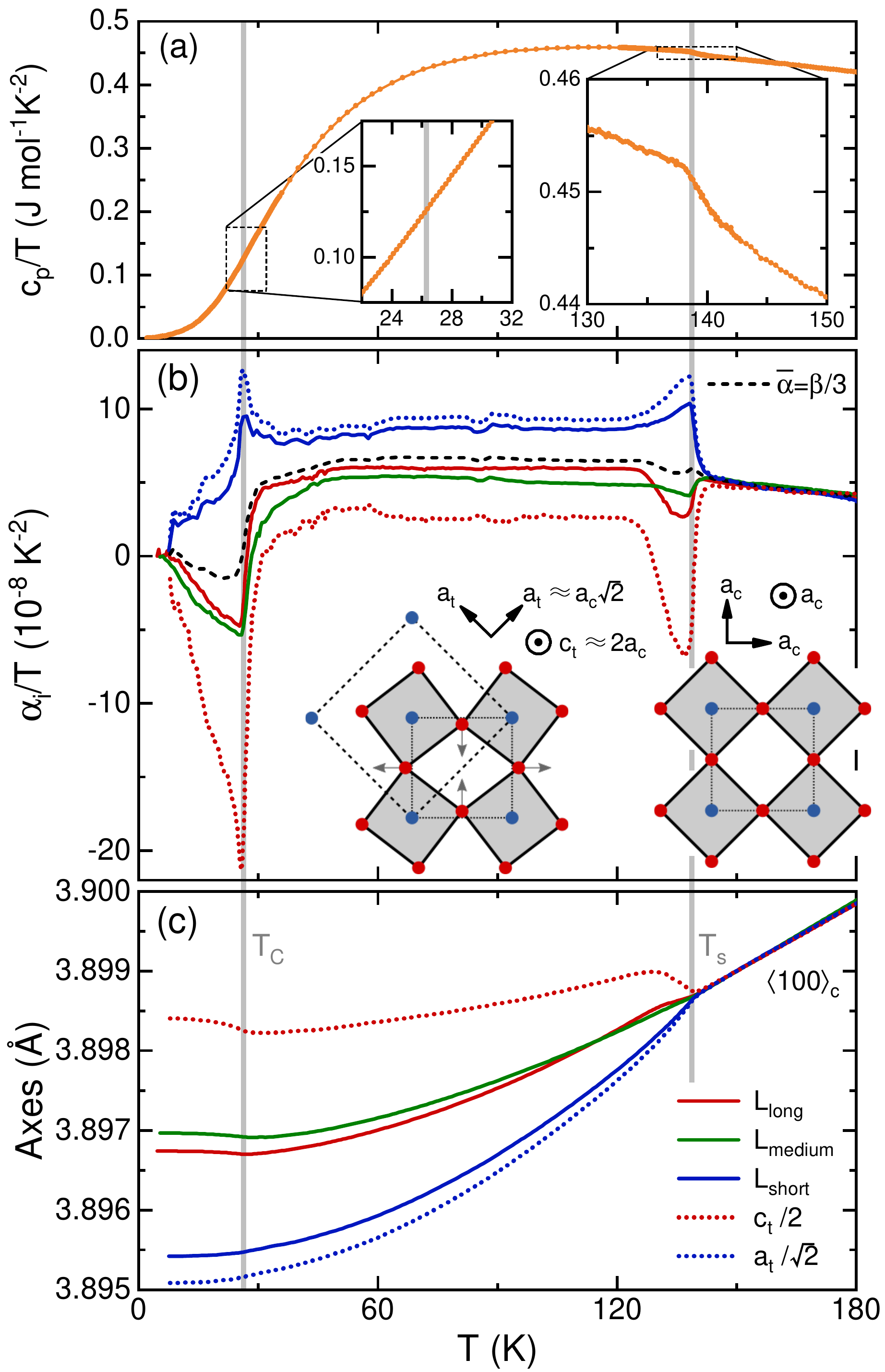}
		\caption{\label{fig:SCTO-anisotropy}
		(a) Specific heat $c_p/T$ of pristine Sr$_{0.991}$Ca$_{0.009}$TiO$_3$. Insets show enlarged views around the structural and ferroelectric transition temperature $T_s$ and $T_\mathrm{C}$, respectively, both indicated by vertical lines.
		(b) Thermal-expansion coefficients $\alpha_i/T$ (thick lines) of pristine Sr$_{0.991}$Ca$_{0.009}$TiO$_3$ measured along the sample's dimensions $L_i$ which are parallel to the cubic axes $\langle 100\rangle$, together with the reconstructed $\alpha_i/T$ of the tetragonal axes $a_\mathrm{t}/\sqrt{2}$ and $c_\mathrm{t}/2$ (dotted lines) and average linear expansion $\bar{\alpha}=\beta/3$ (black dashed line). (c) Corresponding uniaxial length change $\Delta L_i/L_0$ adjusted to the cubic lattice parameter $a_c$ measured at \SI{150}{K} by x-ray diffraction~\cite{Mishra2005}. The sizes and orientations of the cubic and tetragonal unit cells of SrTiO$_3$ are sketched by the insets in (b) with blue~(red) balls representing Ti~(O) ions.}
	\end{figure}

	Figure~\ref{fig:SCTO-anisotropy}~(a) shows the specific heat $c_p/T$ of pristine Sr$_{0.991}$Ca$_{0.009}$TiO$_3$. The insets depict enlarged views of the temperature ranges around both the structural and ferroelectric transition temperature $T_s$ and $T_\mathrm{C}$, respectively, which were determined from thermal-expansion data [see Fig.~\ref{fig:SCTO-anisotropy}~(b) and (c)]. In agreement with previous publications~\cite{DeLima2015,McCalla2016} the structural transition at $T_s$ is clearly visible by a small but distinct $c_p$ anomaly, but no anomaly can be resolved in $c_p(T)$ around $T_\mathrm{C}$.

	Figure~\ref{fig:SCTO-anisotropy}~(b) shows the thermal-expansion coefficients $\alpha_i/T$ of pristine Sr$_{0.991}$Ca$_{0.009}$TiO$_3$ measured along all three of the cubic $\langle 100\rangle$ directions (solid lines) that are parallel to the sample's dimensions $L_i$. Figure~\ref{fig:SCTO-anisotropy}~(c) displays the corresponding uniaxial length changes $\Delta L_i/L_0$ adjusted to the cubic lattice parameter at \SI{150}{K}~\cite{Mishra2005}.
	The $\alpha_i$ are identical at high temperatures and show pronounced anomalies around \SI{139}{K} and \SI{27}{K}. The upper temperature can be identified with the cubic-to-tetragonal transition temperature $T_s$.
	Pure SrTiO$_3$ becomes tetragonal around $T_s\simeq\SI{105}{K}$~\cite{Rimai1962,Lytle1964,Ohama1984,Sato1985} and in Sr$_{1-x}$Ca$_x$TiO$_3$, $T_s$ increases with increasing $x$~\cite{Mitsui1961,Ranjan2001b,Mishra2005,Carpenter2006,DeLima2015}. The transition temperature $T_s\simeq\SI{139}{K}$ of our pristine sample with $x=0.009$ is in agreement with findings in~\cite{Lemanov1997,DeLima2015,McCalla2016}.
	The high-temperature cubic phase has the space group $Pm\bar 3m$~(No. 221) while in the tetragonal phase it is $I4/mcm$~(140)~\cite{Unoki1967}. This transition is antiferrodistortive due to a tilting of the TiO$_6$ octahedra around the $c$ axis~\cite{Unoki1967,Fleury1968,Shirane1969} corresponding to $(a^0a^0c^-)$ in the classification of Glazer~\cite{Glazer1972,Glazer1975}, where $a^0$ denotes the absence of a tilting around the $a$ axis and $c^-$ represents an anti-phase tilting around the $c$ axis. The inset in Fig.~\ref{fig:SCTO-anisotropy}~(b) illustrates the octahedra tilt in a top view of the tetragonal $ab$ plane. Titanium ions (blue) define the corners of the cubic unit cell (dotted square). 
	At $T_s$, the oxygen ions (red) move as indicated by the arrows and consequently, the tetragonal unit cell (dashed square) is doubled in the $ab$ plane and rotated by \ang{45}. Due to the anti-phase tilting in $c$ direction, the $c$ axis is doubled as well. Thus, the cubic axes $a_\mathrm{c}$ and the tetragonal axes $a_\mathrm{t},c_\mathrm{t}$ are related by $a_\mathrm{t}\approx\sqrt{2}a_\mathrm{c}$ and $c_\mathrm{t}\approx 2a_\mathrm{c}$.
	It is evident, that the sample's dimensions $L_i$ which are parallel to the cubic $\langle 100\rangle$ directions, point along $\langle 110\rangle$ with respect to the tetragonal axes $a_\mathrm{t}$.
	The lower transition at \SI{27}{K} signals the ferroelectric phase transition that was characterized by $P(E)$ hysteresis loops~\cite{Rischau2017}.
	While the structural transition at $T_s$ is seen mainly in the sample's short direction $L_\mathrm{short}$, the ferroelectric transition at $T_\mathrm{C}$ predominantly appears along the sample's medium and long direction $L_\mathrm{medium}$ and $L_\mathrm{long}$, respectively.
	
	Structural phase transitions generally involve transformation twinning~\cite{Cahn1954,Authier2003}. For a cubic-to-tetragonal transition one expects the emergence of three twin domains enabling different $\alpha_i$ to partially compensate each other.
	Therefore, a completely twinned sample should exhibit an isotropic uniaxial thermal expansion $\bar\alpha$, which is related to the volume expansion $\beta=3\bar\alpha$. In general, $\beta=\sum_i\alpha_i$, where $\alpha_i$ denote the uniaxial expansion coefficients along a set of three pairwise orthogonal directions and for a tetragonal lattice, $\beta=2\alpha_{a_\mathrm{t}}+\alpha_{c_\mathrm{t}}$ with the (usually anisotropic) main-axis expansion coefficients $\alpha_{a_\mathrm{t}}$ and $\alpha_{c_\mathrm{t}}$ along the tetragonal axes $a_\mathrm{t}$ and $c_\mathrm{t}$, respectively. As shown by $\bar\alpha=\beta/3$ (black dashed line in Fig.~\ref{fig:SCTO-anisotropy}~(b)), the transition
	at $T_s$ is almost volume-conserving, i.e., the expansion in the long direction roughly compensates the contraction along the short and medium direction.
	A nearly volume-conserving transition is also reported for SrTiO$_3$ where the temperature-dependent lattice parameters around $T_s$ were determined by high-resolution x-ray diffraction~\cite{Ohama1984}.
	The anisotropic $\alpha_i$ and the nearly volume conservation indicate highly unequal twinning fractions in our crystal. The usage of a capacitance dilatometer naturally requires the application of a certain uniaxial stress that can be sufficient to achieve a (partial) detwinning of the crystal~\cite{Niesen2013,Kunkemoeller2017}.
	This is apparently not the case in our sample. Here, the presence of a dominating twin is not triggered by external conditions but rather predetermined by intrinsic crystal defects~\footnote{When measuring along $L_\mathrm{long}$, the force applied via the dilatometer acts on the smallest cross-sectional area of our sample and consequently produces the largest pressure, whereas along $L_\mathrm{short}$ the force acts on the largest cross section producing the smallest pressure. Nevertheless, we observe an expansion along $L_\mathrm{long}$ and a compression along $L_\mathrm{short}$ when cooling across $T_s$, which indicates the absence of stress-induced detwinning.}.
	
We can estimate the fraction of the tetragonal axes $a_\mathrm{t}$ and $c_\mathrm{t}$ along the sample's dimensions $L_i$ by comparing the anomalies of the thermal-expansion coefficients of our measurements with the slope changes of the temperature-dependent lattice parameters around $T_s$ from x-ray diffraction measurements. For this comparison, we use the  
high-resolution x-ray data of SrTiO$_3$~\cite{Ohama1984}, which at $T_s$ show an $a_\mathrm{t}$-axis contraction that corresponds to a change $\Delta\alpha_a\simeq\SI{8E-6}{K^{-1}}$ and a $c_\mathrm{t}$-axis expansion corresponding to $\Delta\alpha_c\simeq\SI{-16E-6}{K^{-1}}$. By comparing these values to our thermal-expansion anomalies $\Delta\alpha_i$ at $T_s$ we estimate that the sample's short axis $L_\mathrm{short}$ contains approximately $0.9 a_\mathrm{t}$ and $0.1 c_\mathrm{t}$, whereas the long axis contains $0.5 a_\mathrm{t}+0.5c_\mathrm{t}$, and the medium axis contains $0.6 a_\mathrm{t}+0.4 c_\mathrm{t}$.
From this estimate we reconstruct the tetragonal pure-axes compositions $a_\mathrm{t}=1.25 L_\mathrm{short}-0.25 L_\mathrm{long}$ and $c_\mathrm{t}=2.25L_\mathrm{long}-1.25 L_\mathrm{short}$. Figures \ref{fig:SCTO-anisotropy}~(b) and (c) show the temperature-dependent behavior of the pure tetragonal axes $a_\mathrm{t}/\sqrt{2}$ and $c_\mathrm{t}/2$ as dotted lines.
Of course, the derived anomalies $\Delta\alpha_a$ and $\Delta\alpha_c$ of Sr$_{1-x}$Ca$_x$TiO$_3$ at $T_s$ are by construction identical to those observed by the temperature-dependent x-ray data of the SrTiO$_3$ lattice parameters~\cite{Ohama1984}. The relation $\Delta\alpha_a\approx-\Delta\alpha_c/2$ is, however, independent from this reconstruction and directly follows from the (almost) absent anomaly in the volume expansion $\beta$ and the tetragonal symmetry.
In contrast to the transition at $T_s$, the ferroelectric transition at $T_\mathrm{C}$ is not volume conserving as is clearly demonstrated by the pronounced anomaly in the averaged uniaxial expansion $\bar\alpha$ (see black dashed line in Fig.~\ref{fig:SCTO-anisotropy}~(b)). Furthermore, our reconstructed pure-axis data suggest that the volume-expansion anomaly essentially arises from a $c$-axis expansion upon cooling below $T_\mathrm{C}$, while the transition is roughly area-conserving with respect to the $ab$ plane. The latter is naturally expected for a tetragonal-to-orthorhombic transition with opposite expansion anomalies of similar magnitudes along the orthorhombic $a$ and $b$ axes. Note that the ferroelectric polarization is expected to be aligned along one of these orthorhombic axes, as it has been discussed in~\cite{Bednorz1984,Kleemann1988,Mishra2009}. In this context, it is also worth to mention that a tetragonal-to-orthorhombic transition increases the number of possible twin domains by a factor of two.
However, the uniaxial length change measured along the tetragonal $[110]$ direction cannot distinguish between multi-domain and single-domain orthorhombic samples, because $\alpha_{[110]}=(\alpha_a+\alpha_b)/2$ in both cases.

	\subsection{Electron-doped samples Sr$_{0.991}$Ca$_{0.009}$TiO$_{3-\delta}$}
	\label{doped}

\begin{figure}
	\includegraphics[width=1\columnwidth]{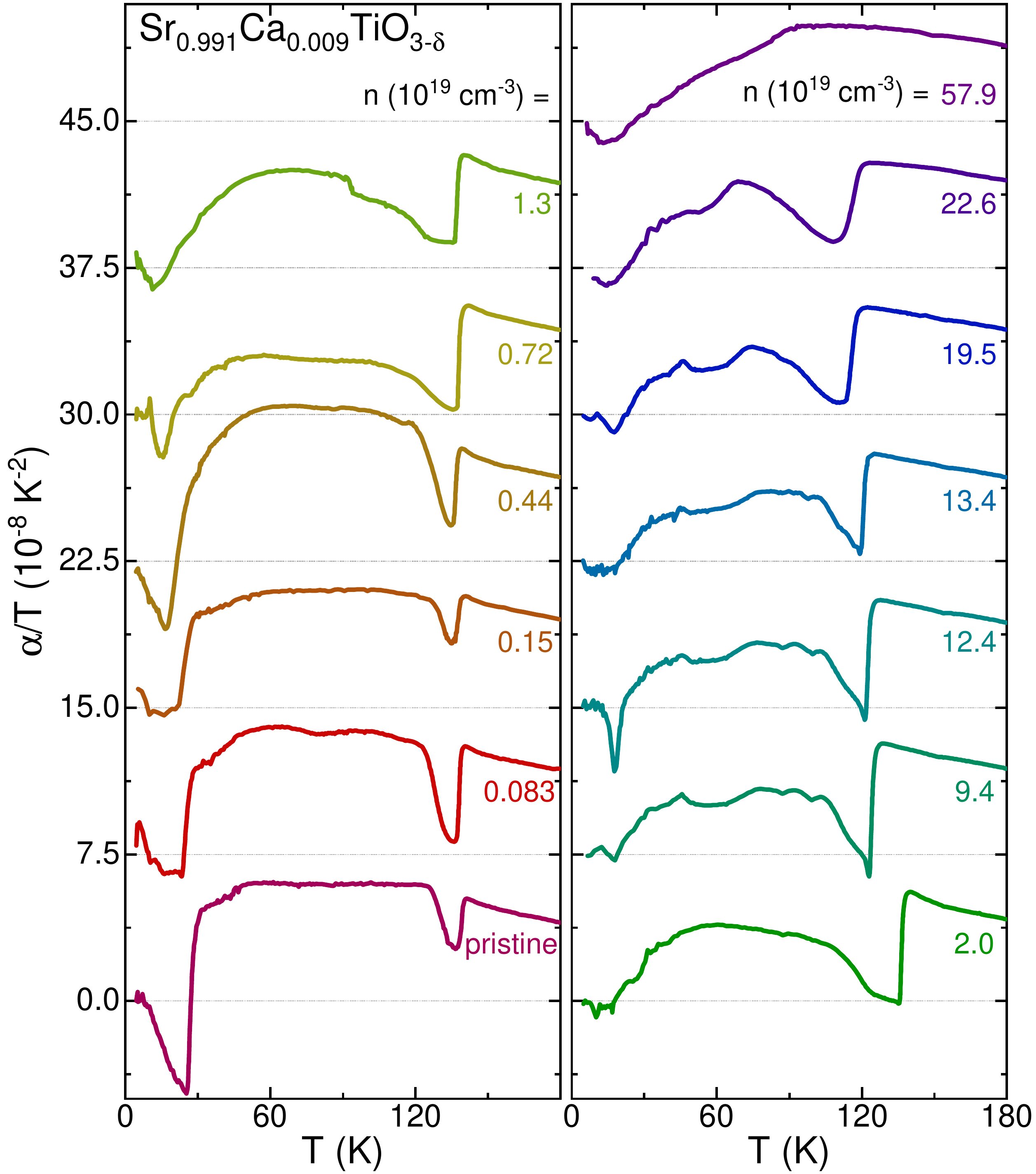}
	\caption{\label{fig:SCTO-alpha-over-T}Thermal-expansion coefficient $\alpha/T$ versus $T$ of Sr$_{0.991}$Ca$_{0.009}$TiO$_{3-\delta}$ with different carrier densities $n$. Curves are shifted with respect to each other.}
\end{figure}

\begin{figure}
	\includegraphics[width=1\columnwidth]{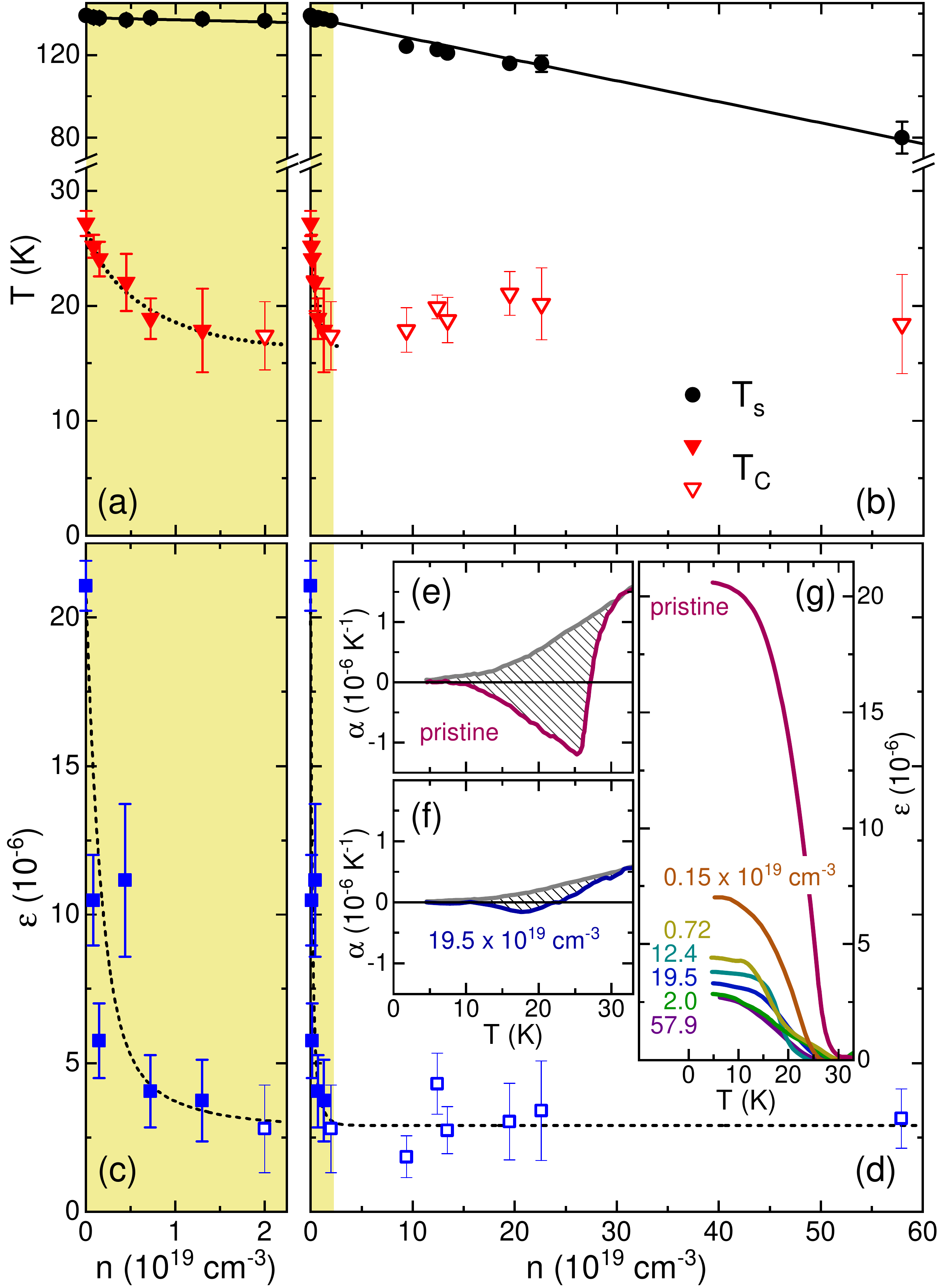}
	\caption{\label{fig:SCTO-phasediag}Phase diagram of Sr$_{0.991}$Ca$_{0.009}$TiO$_{3-\delta}$. Open symbols refer to weak anomalies. (a)~Detail view of the low-$n$ regime with guide to the eye for $T_\mathrm{C}(n)$  (dotted curve). (b)~Complete range of $n$ with linear fit of $T_s(n)$ (black solid line). Note the scale breaks in panels (a) and (b). (c, d) Spontaneous strain $\varepsilon$ as a function of $n$.
		(e, f) We used $\alpha(T)$ of pure SrTiO$_3$ as background (gray curves) and integrated the difference to $\alpha(T)$ of the respective Sr$_{0.991}$Ca$_{0.009}$TiO$_{3-\delta}$ sample, exemplary shown for $n=0$ (e) and $n=\SI{19.5e19}{cm^{-3}}$ (f). (g) Temperature dependence of $\varepsilon$ of selected samples.}
\end{figure}

	Figure~\ref{fig:SCTO-alpha-over-T} shows the thermal-expansion coefficients $\alpha/T$ versus $T$ of Sr$_{0.991}$Ca$_{0.009}$TiO$_{3-\delta}$ with different charge-carrier concentrations up to $n\leq\SI{57.9E19}{cm^{-3}}$ together with $\alpha/T$ of the pristine sample already shown in Fig.~\ref{fig:SCTO-anisotropy}. For clarity, the curves are shifted by \SI{7.5E-8}{K^{-2}} with respect to each other. On each sample, we measured $\alpha_i$ along $L_\mathrm{long}$. Since all samples were obtained by parallel cuts from the original single crystal it appears plausible that the distribution of twin domains does not vary too much over the individual samples. This assumption is essentially confirmed by the fact that all samples, apart from the one with highest $n$, show clear anomalies of the same sign and similar shape signaling the structural phase transition at $T_s$. With increasing $n$ the transition temperature linearly decreases from $T_s\simeq\SI{139}{K}$ in pristine Sr$_{0.991}$Ca$_{0.009}$TiO$_3$ to $\simeq\SI{116}{K}$ for $n=\SI{22.6E19}{cm^{-3}}$ [see Fig.~\ref{fig:SCTO-phasediag}~(b)]. 
	A linear decreasing $T_s(n)$ is known from reduced SrTiO$_{3-\delta}$~\cite{Baeuerle1978a,Wagner1980,Tao2016} and a decreased $T_s$ was also seen in reduced EuTiO$_{3-\delta}$~\cite{Engelmayer2019}, suggesting that this is a generic trend in these almost cubic perovskite titanates~\footnote{The decrease of $T_s$ is, however, not a direct consequence of the charge-carrier doping, because $n$-type doping by chemical substitution like in SrTi$_{1-x}$Nb$_x$O$_3$ increases $T_s$~\cite{Tao2016,McCalla2016}}.
	An extrapolation of the linear $T_s(n)$ dependence to the highest doping $n=\SI{57.9E19}{cm^{-3}}$ matches the kink at  \SI{80}{K} in the $\alpha/T$ curve of the corresponding sample. This suggests that the structural transition remains present in the entire series of Sr$_{0.991}$Ca$_{0.009}$TiO$_{3-\delta}$ samples, but the highest-doped sample apparently has an essentially homogeneous distribution of twin domains and, consequently, the averaged uniaxial expansion $\bar\alpha$ hardly shows any anomaly at this volume-conserving transition as discussed above.

The transition at $T_\mathrm{C}$ is clearly identifiable for the lower-doped samples (left panel of Fig.~\ref{fig:SCTO-alpha-over-T}) and shifts from $T_\mathrm{C}\simeq\SI{27}{K}$ for pristine Sr$_{0.991}$Ca$_{0.009}$TiO$_3$ down to \SI{18}{K} for $n=\SI{1.3E19}{cm^{-3}}$.
For the higher-doped samples (right panel of Fig.~\ref{fig:SCTO-alpha-over-T}) the $\alpha/T$ anomalies become much less pronounced and rather broad; except for the sample with $n=\SI{12.4e19}{cm^{-3}}$. Despite this broadening a signature of the transition remains present in $\alpha/T$ of all doped samples. For example, in the more homogeneously twinned sample with $n=\SI{57.9E19}{cm^{-3}}$ the clear minimum of $\bar\alpha/T$ around \SI{17}{K} signals the spontaneous volume expansion resulting from the low-temperature transition, whereas $\bar\alpha/T$ only shows a kink at $T_s$ of the high-temperature structural transition, which is volume conserving. 

Figure~\ref{fig:SCTO-phasediag} compares the evolution of both transition temperatures $T_s$ and $T_\mathrm{C}$ as a function of charge-carrier density $n$. The linear shift of $T_s$ and the weak broadening of the structural transition with increasing $n$ indicate that both the distribution of oxygen vacancies and the resulting charge-carrier density are rather homogeneous in the studied samples. Nevertheless, the low-temperature transition shows a  complex behavior. The ferroelectric transition of the insulating pristine sample causes a very sharp $\alpha/T$ anomaly, which remains sharp for the low-doped metallic samples but broadens above $n=\SI{1.3E19}{cm^{-3}}$. As a criterion to define $T_\mathrm{C}$, we take the maximum slope of $\alpha/T$ and use the temperature difference to the minimum of $\alpha/T$ as a measure of the transition width, which is shown as error bars in Fig.~\ref{fig:SCTO-phasediag}~(a,b). The corresponding $T_\mathrm{C}(n)$ curve is concave for $n\leq\SI{1.3E19}{cm^{-3}}$, as indicated by the dotted line in Fig.~\ref{fig:SCTO-phasediag}~(a) and essentially saturates for larger $n$. As a further measure of this transition, we use a smooth background $\alpha_\mathrm{bg}$ and calculate $\varepsilon=\int(\alpha-\alpha_\mathrm{bg})\,\mathrm{d}T$, which yields the spontaneous elongation resulting from the low-temperature transition. For $\alpha_\mathrm{bg}(T)$ we measured the thermal expansion $\alpha_\mathrm{STO}$ on a single crystal of SrTiO$_3$~\footnote{On our twinned SrTiO$_3$ crystal we find a smooth $\alpha(T)$ low-temperature behavior. In contrast, thermal-expansion data on SrTiO$_3$ measured under strong uniaxial compressive stress result in single-domain tetragonal samples with very different expansion~\cite{Tsunekawa1984}. The corresponding $\alpha(T)$ curves show a minimum between \SI{20}{K} and \SI{30}{K}, which resembles the data of our Ca-doped samples. Because SrTiO$_3$ is known to develop a stress-induced ferroelectric order~\cite{Uwe1976}, the anomalous $\alpha(T)$ curves observed in~\cite{Tsunekawa1984} probably arise from ferroelectric order that is induced in those SrTiO$_3$ crystals by the application of the uniaxial compressive stress.}, which remains in the tetragonal phase and scaled this $\alpha_\mathrm{STO}(T)$ curve such that it matches $\alpha(T\approx\SI{25}{K})$ of the respective Sr$_{0.991}$Ca$_{0.009}$TiO$_{3-\delta}$ sample. This is shown for 2 exemplary charge-carrier contents in Fig.~\ref{fig:SCTO-phasediag}~(e,f) and the resulting $\varepsilon(T)$ are displayed in Fig.~\ref{fig:SCTO-phasediag}~(g) with every second measurement skipped for clarity. The evolution of $\varepsilon(T=\SI{4.2}{K},n)$ as a function $n$ is summarized in Fig.~\ref{fig:SCTO-phasediag}~(c,d); the corresponding error bars express the sensitivity of  $\varepsilon(T=\SI{4.2}{K},n)$ on variations of the individual scaling factors used for $\alpha_\mathrm{STO}$. We see that $\varepsilon(T=\SI{4.2}{K},n)$ rapidly decreases at $n\leq\SI{1.3E19}{cm^{-3}}$ and then levels off at a value of about 15\,\% of the spontaneous elongation measured in the ferroelectric phase of pristine Sr$_{0.991}$Ca$_{0.009}$TiO$_{3}$. This behavior is qualitatively similar to that of $T_\mathrm{C}(n)$, which saturates, however, at a comparatively larger value of about 60\,\% of  $T_\mathrm{C}$ of the insulating pristine material.    

These are the main findings of the present study, which extends our previous work suggesting the existence of a ferroelectric quantum phase transition inside the superconducting dome of Sr$_{1-x}$Ca$_x$TiO$_{3-\delta}$~\cite{Rischau2017}. As already discussed there, a true ferroelectric order cannot exist in metals, because the mobile electrons screen any static electric polarization. 
However, characteristic features of the ferroelectric order in Sr$_{1-x}$Ca$_x$TiO$_3$ are still observed upon weak charge-carrier doping suggesting a ferroelectriclike transition that vanishes via a quantum phase transition as indicated by minima in the resitivity data $\rho(T,n)$~\cite{Rischau2017}.
The aim of the present study was to further clarify this issue via thermal-expansion measurements, which are a sensitive thermodynamic probe to detect and characterize pressure-dependent quantum phase transitions~\cite{Zhu2003,Garst2005,Kuechler2003,Gegenwart2006,Baier2007,Lorenz2007,Lorenz2008,Breunig2013,Breunig2017a,Grube2018,Meingast2012}.
The existence of a well-defined ferroelectriclike transition in the metallic samples is clearly confirmed by the pronounced $\alpha/T$ anomalies (see Fig.~\ref{fig:SCTO-alpha-over-T}) up to at least $n=\SI{0.72e19}{cm^{-3}}$, i.e., the transition remains present in metallic samples which finally become superconducting at lower temperature~\cite{Rischau2017}. On further increasing $n$, however, the $\alpha/T$ anomalies do not vanish and, in particular, our data do not give any indication for a sign change of the $\alpha/T$ anomalies as a function of $n$. Thus, the general behavior of $\alpha$ and the corresponding Gr\"{u}neisen ratio $\Gamma=\alpha/c_p$ is different from other materials showing quantum phase transitions as a function of magnetic field, pressure, or $n$ as external control parameter~\cite{Kuechler2003,Gegenwart2006,Baier2007,Lorenz2007,Lorenz2008,Breunig2013,Breunig2017a,Grube2018,Meingast2012}.
On the one hand, this could mean the absence of a quantum phase transition, if the ferroelectriclike order changes just continuously disappears towards larger $n$.
On the other hand, however, it is also clear from Figs.~\ref{fig:SCTO-alpha-over-T} and~\ref{fig:SCTO-phasediag} that the shape and width of the $\alpha/T$ anomalies strongly change around $n=\SI{1.3e19}{cm^{-3}}$.
This could mean that the symmetry change of the structural transition is different above and below a critical doping $n^\star$ in the range of $\SI{1.3e19}{cm^{-3}}$.
A similar situation is, in fact, observed in the undoped Sr$_{1-x}$Ca$_x$TiO$_3$ where, as a function of $x$, the symmetry of the low-temperature ordered phase changes from polar ferroelectric (space group $Ic2m$; No.~46) via relaxor ferroelectric to antiferroelectric with inversion symmetry ($Pbcm$, No.~57) above $x\simeq 0.12$~\cite{Bednorz1984,Ranjan2000,Ranjan2001a,Ranjan2001b,Mishra2002}.
Such a microscopic change is not necessarily reflected in the thermal-expansion coefficient.
In the simplest case, this could be an experimental problem, because a more or less vertical phase boundary in an $n$-$T$ phase diagram is difficult to measure as a function of $T$. More generally, anomalies in the uniaxial expansion $\alpha$ require a finite dependence of the respective transition temperature on uniaxial stress along this direction~\footnote{This follows from general thermodynamics via Clausius-Clapeyron (Ehrenfest) relations for first- (second-)order phase transitions}.
In this context, it is also important that the polarization of Sr$_{1-x}$Ca$_x$TiO$_3$ is in the orthorhombic $ab$ plane~\cite{Bednorz1984,Kleemann1988,Mishra2009}, such that a change of the in-plane symmetry will have minor influence on the observed $\alpha/T$ anomaly, because it arises from a spontaneous elongation of the $c$-axis component in our partially twinned crystals, as discussed above in the context of Fig.~\ref{fig:SCTO-anisotropy}. Thus, detailed structural analysis of the low-temperature phases for different charge-carrier concentrations would be necessary to resolve this puzzle.

\section{Summary}

In conclusion, we present a detailed thermal-expansion study on Sr$_{1-x}$Ca$_x$TiO$_{3-\delta}$ ($x=0.009$) with charge-carrier density tuned from the pristine, insulating parent compound to $n\simeq\SI{60e19}{cm^{-3}}$. Both the cubic-to-tetragonal transition $T_s$ and the ferroelectric transition $T_\mathrm{C}$ display pronounced anomalies in the thermal-expansion coefficient $\alpha(T)$ of the pristine crystal.
As a function of charge-carrier density $n$, $T_s$ decreases linearly from \SI{139}{K} to \SI{80}{K} and the related anomalies in $\alpha/T$ remain distinct and of similar magnitude across almost the entire doping range.
Despite the presence of mobile charge carriers, the $T_\mathrm{C}$-related anomaly survives for $n>0$. However, $T_\mathrm{C}$ decreases rapidly upon increasing $n$ and the associated anomalies in $\alpha/T$ become very broad above $n\simeq\SI{1.3e19}{cm^{-3}}$. Whether the evolution of $T_\mathrm{C}$ across this carrier concentration is continuous or passes a phase transition at a critical $n^\star$ is not directly seen in $\alpha(T)$ but needs to be clarified by a structure analysis.

	\begin{acknowledgments}
		We acknowledge support by the DFG (German Research Foundation) via project number 277146847 - CRC 1238 (Subprojects A02, B01 and B02). This work is part of a DFG-ANR project funded by Agence Nationale de la Recherche (ANR-18-CE92-0020-01) and by the DFG through projects LO~818/6-1 and HE~3219/6-1. X.~L. acknowledges support by the Alexander von Humboldt Foundation and Zhejiang Provincial Natural Science Foundation of China under Grant No. LQ19A040005.
	\end{acknowledgments}


\begin{thebibliography}{73}%
\makeatletter
\providecommand \@ifxundefined [1]{%
 \@ifx{#1\undefined}
}%
\providecommand \@ifnum [1]{%
 \ifnum #1\expandafter \@firstoftwo
 \else \expandafter \@secondoftwo
 \fi
}%
\providecommand \@ifx [1]{%
 \ifx #1\expandafter \@firstoftwo
 \else \expandafter \@secondoftwo
 \fi
}%
\providecommand \natexlab [1]{#1}%
\providecommand \enquote  [1]{``#1''}%
\providecommand \bibnamefont  [1]{#1}%
\providecommand \bibfnamefont [1]{#1}%
\providecommand \citenamefont [1]{#1}%
\providecommand \href@noop [0]{\@secondoftwo}%
\providecommand \href [0]{\begingroup \@sanitize@url \@href}%
\providecommand \@href[1]{\@@startlink{#1}\@@href}%
\providecommand \@@href[1]{\endgroup#1\@@endlink}%
\providecommand \@sanitize@url [0]{\catcode `\\12\catcode `\$12\catcode
  `\&12\catcode `\#12\catcode `\^12\catcode `\_12\catcode `\%12\relax}%
\providecommand \@@startlink[1]{}%
\providecommand \@@endlink[0]{}%
\providecommand \url  [0]{\begingroup\@sanitize@url \@url }%
\providecommand \@url [1]{\endgroup\@href {#1}{\urlprefix }}%
\providecommand \urlprefix  [0]{URL }%
\providecommand \Eprint [0]{\href }%
\providecommand \doibase [0]{https://doi.org/}%
\providecommand \selectlanguage [0]{\@gobble}%
\providecommand \bibinfo  [0]{\@secondoftwo}%
\providecommand \bibfield  [0]{\@secondoftwo}%
\providecommand \translation [1]{[#1]}%
\providecommand \BibitemOpen [0]{}%
\providecommand \bibitemStop [0]{}%
\providecommand \bibitemNoStop [0]{.\EOS\space}%
\providecommand \EOS [0]{\spacefactor3000\relax}%
\providecommand \BibitemShut  [1]{\csname bibitem#1\endcsname}%
\let\auto@bib@innerbib\@empty
\bibitem [{\citenamefont {{von Hippel}}(1950)}]{VonHippel1950}%
  \BibitemOpen
  \bibfield  {author} {\bibinfo {author} {\bibfnamefont {A.}~\bibnamefont {{von
  Hippel}}},\ }\bibfield  {title} {\bibinfo {title} {{Ferroelectricity, Domain
  Structure, and Phase Transitions of Barium Titanate}},\ }\href
  {https://doi.org/10.1103/revmodphys.22.221} {\bibfield  {journal} {\bibinfo
  {journal} {Rev. Mod. Phys.}\ }\textbf {\bibinfo {volume} {22}},\ \bibinfo
  {pages} {221} (\bibinfo {year} {1950})}\BibitemShut {NoStop}%
\bibitem [{\citenamefont {Bhide}\ \emph {et~al.}(1962)\citenamefont {Bhide},
  \citenamefont {Deshmukh},\ and\ \citenamefont {Hegde}}]{Bhide1962}%
  \BibitemOpen
  \bibfield  {author} {\bibinfo {author} {\bibfnamefont {V.~G.}\ \bibnamefont
  {Bhide}}, \bibinfo {author} {\bibfnamefont {K.~G.}\ \bibnamefont
  {Deshmukh}},\ and\ \bibinfo {author} {\bibfnamefont {M.~S.}\ \bibnamefont
  {Hegde}},\ }\bibfield  {title} {\bibinfo {title} {{Ferroelectric properties
  of PbTiO$_3$}},\ }\href {https://doi.org/10.1016/0031-8914(62)90075-7}
  {\bibfield  {journal} {\bibinfo  {journal} {Physica}\ }\textbf {\bibinfo
  {volume} {28}},\ \bibinfo {pages} {871} (\bibinfo {year} {1962})}\BibitemShut
  {NoStop}%
\bibitem [{\citenamefont {Sun}\ \emph {et~al.}(1998)\citenamefont {Sun},
  \citenamefont {Nakamura}, \citenamefont {Shan}, \citenamefont {Inaguma},\
  and\ \citenamefont {Itoh}}]{Sun1998}%
  \BibitemOpen
  \bibfield  {author} {\bibinfo {author} {\bibfnamefont {P.-H.}\ \bibnamefont
  {Sun}}, \bibinfo {author} {\bibfnamefont {T.}~\bibnamefont {Nakamura}},
  \bibinfo {author} {\bibfnamefont {Y.~J.}\ \bibnamefont {Shan}}, \bibinfo
  {author} {\bibfnamefont {Y.}~\bibnamefont {Inaguma}},\ and\ \bibinfo {author}
  {\bibfnamefont {M.}~\bibnamefont {Itoh}},\ }\bibfield  {title} {\bibinfo
  {title} {{The study on the dielectric property and structure of perovskite
  titanate CdTiO$_3$}},\ }\href {https://doi.org/10.1080/00150199808015031}
  {\bibfield  {journal} {\bibinfo  {journal} {Ferroelectrics}\ }\textbf
  {\bibinfo {volume} {217}},\ \bibinfo {pages} {137} (\bibinfo {year}
  {1998})}\BibitemShut {NoStop}%
\bibitem [{\citenamefont {Kennedy}\ \emph {et~al.}(2011)\citenamefont
  {Kennedy}, \citenamefont {Zhou},\ and\ \citenamefont {Avdeev}}]{Kennedy2011}%
  \BibitemOpen
  \bibfield  {author} {\bibinfo {author} {\bibfnamefont {B.~J.}\ \bibnamefont
  {Kennedy}}, \bibinfo {author} {\bibfnamefont {Q.}~\bibnamefont {Zhou}},\ and\
  \bibinfo {author} {\bibfnamefont {M.}~\bibnamefont {Avdeev}},\ }\bibfield
  {title} {\bibinfo {title} {{The ferroelectric phase of CdTiO$_3$: A powder
  neutron diffraction study}},\ }\href
  {https://doi.org/10.1016/j.jssc.2011.08.028} {\bibfield  {journal} {\bibinfo
  {journal} {J. Solid State Chem.}\ }\textbf {\bibinfo {volume} {184}},\
  \bibinfo {pages} {2987} (\bibinfo {year} {2011})}\BibitemShut {NoStop}%
\bibitem [{\citenamefont {Nakamura}\ \emph {et~al.}(1997)\citenamefont
  {Nakamura}, \citenamefont {Sun}, \citenamefont {Shan}, \citenamefont
  {Inaguma}, \citenamefont {Itoh}, \citenamefont {Kim}, \citenamefont {Sohn},
  \citenamefont {Ikeda}, \citenamefont {Kitamura},\ and\ \citenamefont
  {Konagaya}}]{Nakamura1997}%
  \BibitemOpen
  \bibfield  {author} {\bibinfo {author} {\bibfnamefont {T.}~\bibnamefont
  {Nakamura}}, \bibinfo {author} {\bibfnamefont {P.-H.}\ \bibnamefont {Sun}},
  \bibinfo {author} {\bibfnamefont {Y.~J.}\ \bibnamefont {Shan}}, \bibinfo
  {author} {\bibfnamefont {Y.}~\bibnamefont {Inaguma}}, \bibinfo {author}
  {\bibfnamefont {M.}~\bibnamefont {Itoh}}, \bibinfo {author} {\bibfnamefont
  {I.-S.}\ \bibnamefont {Kim}}, \bibinfo {author} {\bibfnamefont {J.-H.}\
  \bibnamefont {Sohn}}, \bibinfo {author} {\bibfnamefont {M.}~\bibnamefont
  {Ikeda}}, \bibinfo {author} {\bibfnamefont {T.}~\bibnamefont {Kitamura}},\
  and\ \bibinfo {author} {\bibfnamefont {H.}~\bibnamefont {Konagaya}},\
  }\bibfield  {title} {\bibinfo {title} {{On the perovskite-related materials
  of high dielectric permittivity with small temperature dependence and low
  dielectric loss}},\ }\href {https://doi.org/10.1080/00150199708224163}
  {\bibfield  {journal} {\bibinfo  {journal} {Ferroelectrics}\ }\textbf
  {\bibinfo {volume} {196}},\ \bibinfo {pages} {205} (\bibinfo {year}
  {1997})}\BibitemShut {NoStop}%
\bibitem [{\citenamefont {Kim}\ \emph {et~al.}(1992)\citenamefont {Kim},
  \citenamefont {Itoh},\ and\ \citenamefont {Nakamura}}]{Kim1992}%
  \BibitemOpen
  \bibfield  {author} {\bibinfo {author} {\bibfnamefont {I.-S.}\ \bibnamefont
  {Kim}}, \bibinfo {author} {\bibfnamefont {M.}~\bibnamefont {Itoh}},\ and\
  \bibinfo {author} {\bibfnamefont {T.}~\bibnamefont {Nakamura}},\ }\bibfield
  {title} {\bibinfo {title} {{Electrical conductivity and metal-nonmetal
  transition in the perovskite-related layered system
  Ca$_{n+1}$Ti$_n$O$_{3n+1-\delta}$ ($n=2,3,\infty$)}},\ }\href
  {https://doi.org/10.1016/0022-4596(92)90203-8} {\bibfield  {journal}
  {\bibinfo  {journal} {J. Solid State Chem.}\ }\textbf {\bibinfo {volume}
  {101}},\ \bibinfo {pages} {77} (\bibinfo {year} {1992})}\BibitemShut
  {NoStop}%
\bibitem [{\citenamefont {Lemanov}\ \emph {et~al.}(1999)\citenamefont
  {Lemanov}, \citenamefont {Sotnikov}, \citenamefont {Smirnova}, \citenamefont
  {Weihnacht},\ and\ \citenamefont {Kunze}}]{Lemanov1999}%
  \BibitemOpen
  \bibfield  {author} {\bibinfo {author} {\bibfnamefont {V.~V.}\ \bibnamefont
  {Lemanov}}, \bibinfo {author} {\bibfnamefont {A.~V.}\ \bibnamefont
  {Sotnikov}}, \bibinfo {author} {\bibfnamefont {E.~P.}\ \bibnamefont
  {Smirnova}}, \bibinfo {author} {\bibfnamefont {M.}~\bibnamefont
  {Weihnacht}},\ and\ \bibinfo {author} {\bibfnamefont {R.}~\bibnamefont
  {Kunze}},\ }\bibfield  {title} {\bibinfo {title} {{Perovskite CaTiO$_3$ as an
  incipient ferroelectric}},\ }\href
  {https://doi.org/10.1016/S0038-1098(99)00153-2} {\bibfield  {journal}
  {\bibinfo  {journal} {Solid State Commun.}\ }\textbf {\bibinfo {volume}
  {110}},\ \bibinfo {pages} {611} (\bibinfo {year} {1999})}\BibitemShut
  {NoStop}%
\bibitem [{\citenamefont {M{\"{u}}ller}\ and\ \citenamefont
  {Burkard}(1979)}]{Mueller1979}%
  \BibitemOpen
  \bibfield  {author} {\bibinfo {author} {\bibfnamefont {K.~A.}\ \bibnamefont
  {M{\"{u}}ller}}\ and\ \bibinfo {author} {\bibfnamefont {H.}~\bibnamefont
  {Burkard}},\ }\bibfield  {title} {\bibinfo {title} {{SrTiO$_3$: An intrinsic
  quantum paraelectric below 4 K}},\ }\href
  {https://doi.org/10.1103/PhysRevB.19.3593} {\bibfield  {journal} {\bibinfo
  {journal} {Phys. Rev. B}\ }\textbf {\bibinfo {volume} {19}},\ \bibinfo
  {pages} {3593} (\bibinfo {year} {1979})}\BibitemShut {NoStop}%
\bibitem [{\citenamefont {Katsufuji}\ and\ \citenamefont
  {Takagi}(2001)}]{Katsufuji2001}%
  \BibitemOpen
  \bibfield  {author} {\bibinfo {author} {\bibfnamefont {T.}~\bibnamefont
  {Katsufuji}}\ and\ \bibinfo {author} {\bibfnamefont {H.}~\bibnamefont
  {Takagi}},\ }\bibfield  {title} {\bibinfo {title} {{Coupling between
  magnetism and dielectric properties in quantum paraelectric EuTiO$_3$}},\
  }\href {https://doi.org/10.1103/PhysRevB.64.054415} {\bibfield  {journal}
  {\bibinfo  {journal} {Phys. Rev. B}\ }\textbf {\bibinfo {volume} {64}},\
  \bibinfo {pages} {054415} (\bibinfo {year} {2001})}\BibitemShut {NoStop}%
\bibitem [{\citenamefont {Kamba}\ \emph {et~al.}(2007)\citenamefont {Kamba},
  \citenamefont {Nuzhnyy}, \citenamefont {Van{\v{e}}k}, \citenamefont
  {Savinov}, \citenamefont {Kn{\'{i}}{\v{z}}ek}, \citenamefont {Shen},
  \citenamefont {{\v{S}}antav{\'{a}}}, \citenamefont {Maca}, \citenamefont
  {Sadowski},\ and\ \citenamefont {Petzelt}}]{Kamba2007}%
  \BibitemOpen
  \bibfield  {author} {\bibinfo {author} {\bibfnamefont {S.}~\bibnamefont
  {Kamba}}, \bibinfo {author} {\bibfnamefont {D.}~\bibnamefont {Nuzhnyy}},
  \bibinfo {author} {\bibfnamefont {P.}~\bibnamefont {Van{\v{e}}k}}, \bibinfo
  {author} {\bibfnamefont {M.}~\bibnamefont {Savinov}}, \bibinfo {author}
  {\bibfnamefont {K.}~\bibnamefont {Kn{\'{i}}{\v{z}}ek}}, \bibinfo {author}
  {\bibfnamefont {Z.}~\bibnamefont {Shen}}, \bibinfo {author} {\bibfnamefont
  {E.}~\bibnamefont {{\v{S}}antav{\'{a}}}}, \bibinfo {author} {\bibfnamefont
  {K.}~\bibnamefont {Maca}}, \bibinfo {author} {\bibfnamefont {M.}~\bibnamefont
  {Sadowski}},\ and\ \bibinfo {author} {\bibfnamefont {J.}~\bibnamefont
  {Petzelt}},\ }\bibfield  {title} {\bibinfo {title} {{Magnetodielectric effect
  and optic soft mode behaviour in quantum paraelectric EuTiO$_3$ ceramics}},\
  }\href {https://doi.org/10.1209/0295-5075/80/27002} {\bibfield  {journal}
  {\bibinfo  {journal} {Europhys. Lett.}\ }\textbf {\bibinfo {volume} {80}},\
  \bibinfo {pages} {27002} (\bibinfo {year} {2007})}\BibitemShut {NoStop}%
\bibitem [{\citenamefont {Engelmayer}\ \emph {et~al.}(2019)\citenamefont
  {Engelmayer}, \citenamefont {Lin}, \citenamefont {Grams}, \citenamefont
  {German}, \citenamefont {Fr{\"{o}}hlich}, \citenamefont {Hemberger},
  \citenamefont {Behnia},\ and\ \citenamefont {Lorenz}}]{Engelmayer2019}%
  \BibitemOpen
  \bibfield  {author} {\bibinfo {author} {\bibfnamefont {J.}~\bibnamefont
  {Engelmayer}}, \bibinfo {author} {\bibfnamefont {X.}~\bibnamefont {Lin}},
  \bibinfo {author} {\bibfnamefont {C.~P.}\ \bibnamefont {Grams}}, \bibinfo
  {author} {\bibfnamefont {R.}~\bibnamefont {German}}, \bibinfo {author}
  {\bibfnamefont {T.}~\bibnamefont {Fr{\"{o}}hlich}}, \bibinfo {author}
  {\bibfnamefont {J.}~\bibnamefont {Hemberger}}, \bibinfo {author}
  {\bibfnamefont {K.}~\bibnamefont {Behnia}},\ and\ \bibinfo {author}
  {\bibfnamefont {T.}~\bibnamefont {Lorenz}},\ }\bibfield  {title} {\bibinfo
  {title} {{Charge transport in oxygen-deficient EuTiO$_3$: The emerging
  picture of dilute metallicity in quantum-paraelectric perovskite oxides}},\
  }\href {https://doi.org/10.1103/physrevmaterials.3.051401} {\bibfield
  {journal} {\bibinfo  {journal} {Phys. Rev. Materials}\ }\textbf {\bibinfo
  {volume} {3}},\ \bibinfo {pages} {051401(R)} (\bibinfo {year}
  {2019})}\BibitemShut {NoStop}%
\bibitem [{\citenamefont {Hulm}(1950)}]{Hulm1950}%
  \BibitemOpen
  \bibfield  {author} {\bibinfo {author} {\bibfnamefont {J.~K.}\ \bibnamefont
  {Hulm}},\ }\bibfield  {title} {\bibinfo {title} {{The Dielectric Properties
  of some Alkaline Earth Titanates at Low Temperatures}},\ }\href
  {https://doi.org/10.1088/0370-1298/63/10/118} {\bibfield  {journal} {\bibinfo
   {journal} {Proc. Phys. Soc. London, Sect. A}\ }\textbf {\bibinfo {volume}
  {63}},\ \bibinfo {pages} {1184} (\bibinfo {year} {1950})}\BibitemShut
  {NoStop}%
\bibitem [{\citenamefont {Bednorz}\ and\ \citenamefont
  {M{\"{u}}ller}(1984)}]{Bednorz1984}%
  \BibitemOpen
  \bibfield  {author} {\bibinfo {author} {\bibfnamefont {J.~G.}\ \bibnamefont
  {Bednorz}}\ and\ \bibinfo {author} {\bibfnamefont {K.~A.}\ \bibnamefont
  {M{\"{u}}ller}},\ }\bibfield  {title} {\bibinfo {title}
  {{Sr$_{1-x}$Ca$_x$TiO$_3$: An $XY$ Quantum Ferroelectric with Transition to
  Randomness}},\ }\href {https://doi.org/10.1103/PhysRevLett.52.2289}
  {\bibfield  {journal} {\bibinfo  {journal} {Phys. Rev. Lett.}\ }\textbf
  {\bibinfo {volume} {52}},\ \bibinfo {pages} {2289} (\bibinfo {year}
  {1984})}\BibitemShut {NoStop}%
\bibitem [{\citenamefont {Ranjan}\ \emph {et~al.}(2000)\citenamefont {Ranjan},
  \citenamefont {Pandey},\ and\ \citenamefont {Lalla}}]{Ranjan2000}%
  \BibitemOpen
  \bibfield  {author} {\bibinfo {author} {\bibfnamefont {R.}~\bibnamefont
  {Ranjan}}, \bibinfo {author} {\bibfnamefont {D.}~\bibnamefont {Pandey}},\
  and\ \bibinfo {author} {\bibfnamefont {N.~P.}\ \bibnamefont {Lalla}},\
  }\bibfield  {title} {\bibinfo {title} {{Novel Features of
  Sr$_{1-x}$Ca$_x$TiO$_3$ Phase Diagram: Evidence for Competing
  Antiferroelectric and Ferroelectric Interactions}},\ }\href
  {https://doi.org/10.1103/physrevlett.84.3726} {\bibfield  {journal} {\bibinfo
   {journal} {Phys. Rev. Lett.}\ }\textbf {\bibinfo {volume} {84}},\ \bibinfo
  {pages} {3726} (\bibinfo {year} {2000})}\BibitemShut {NoStop}%
\bibitem [{\citenamefont {Ranjan}\ and\ \citenamefont
  {Pandey}(2001{\natexlab{a}})}]{Ranjan2001a}%
  \BibitemOpen
  \bibfield  {author} {\bibinfo {author} {\bibfnamefont {R.}~\bibnamefont
  {Ranjan}}\ and\ \bibinfo {author} {\bibfnamefont {D.}~\bibnamefont
  {Pandey}},\ }\bibfield  {title} {\bibinfo {title} {{Antiferroelectric phase
  transition in (Sr$_{1-x}$Ca$_x$)TiO$_3$ ($0.12<x\leq 0.40$): I. Dielectric
  studies}},\ }\href {https://doi.org/10.1088/0953-8984/13/19/305} {\bibfield
  {journal} {\bibinfo  {journal} {J. Phys.: Condens. Matter}\ }\textbf
  {\bibinfo {volume} {13}},\ \bibinfo {pages} {4239} (\bibinfo {year}
  {2001}{\natexlab{a}})}\BibitemShut {NoStop}%
\bibitem [{\citenamefont {Ranjan}\ and\ \citenamefont
  {Pandey}(2001{\natexlab{b}})}]{Ranjan2001b}%
  \BibitemOpen
  \bibfield  {author} {\bibinfo {author} {\bibfnamefont {R.}~\bibnamefont
  {Ranjan}}\ and\ \bibinfo {author} {\bibfnamefont {D.}~\bibnamefont
  {Pandey}},\ }\bibfield  {title} {\bibinfo {title} {{Antiferroelectric phase
  transition in (Sr$_{1-x}$Ca$_x$)TiO$_3$: II. X-ray diffraction studies}},\
  }\href {https://doi.org/10.1088/0953-8984/13/19/306} {\bibfield  {journal}
  {\bibinfo  {journal} {J. Phys.: Condens. Matter}\ }\textbf {\bibinfo {volume}
  {13}},\ \bibinfo {pages} {4251} (\bibinfo {year}
  {2001}{\natexlab{b}})}\BibitemShut {NoStop}%
\bibitem [{\citenamefont {Mishra}\ \emph {et~al.}(2002)\citenamefont {Mishra},
  \citenamefont {Ranjan}, \citenamefont {Pandey},\ and\ \citenamefont
  {Kennedy}}]{Mishra2002}%
  \BibitemOpen
  \bibfield  {author} {\bibinfo {author} {\bibfnamefont {S.~K.}\ \bibnamefont
  {Mishra}}, \bibinfo {author} {\bibfnamefont {R.}~\bibnamefont {Ranjan}},
  \bibinfo {author} {\bibfnamefont {D.}~\bibnamefont {Pandey}},\ and\ \bibinfo
  {author} {\bibfnamefont {B.~J.}\ \bibnamefont {Kennedy}},\ }\bibfield
  {title} {\bibinfo {title} {{Powder neutron diffraction study of the
  antiferroelectric phase transition in Sr$_{0.75}$Ca$_{0.25}$TiO$_3$}},\
  }\href {https://doi.org/10.1063/1.1432475} {\bibfield  {journal} {\bibinfo
  {journal} {J. Appl. Phys.}\ }\textbf {\bibinfo {volume} {91}},\ \bibinfo
  {pages} {4447} (\bibinfo {year} {2002})}\BibitemShut {NoStop}%
\bibitem [{\citenamefont {Bianchi}\ \emph {et~al.}(1994)\citenamefont
  {Bianchi}, \citenamefont {Kleemann},\ and\ \citenamefont
  {Bednorz}}]{Bianchi1994}%
  \BibitemOpen
  \bibfield  {author} {\bibinfo {author} {\bibfnamefont {U.}~\bibnamefont
  {Bianchi}}, \bibinfo {author} {\bibfnamefont {W.}~\bibnamefont {Kleemann}},\
  and\ \bibinfo {author} {\bibfnamefont {J.~G.}\ \bibnamefont {Bednorz}},\
  }\bibfield  {title} {\bibinfo {title} {{Raman scattering of ferroelectric
  Sr$_{1-x}$Ca$_x$TiO$_3$, $x=0.007$}},\ }\href
  {https://doi.org/10.1088/0953-8984/6/6/025} {\bibfield  {journal} {\bibinfo
  {journal} {J. Phys.: Condens. Matter}\ }\textbf {\bibinfo {volume} {6}},\
  \bibinfo {pages} {1229} (\bibinfo {year} {1994})}\BibitemShut {NoStop}%
\bibitem [{\citenamefont {Kleemann}\ \emph {et~al.}(1997)\citenamefont
  {Kleemann}, \citenamefont {Albertini}, \citenamefont {Kuss},\ and\
  \citenamefont {Lindner}}]{Kleemann1997}%
  \BibitemOpen
  \bibfield  {author} {\bibinfo {author} {\bibfnamefont {W.}~\bibnamefont
  {Kleemann}}, \bibinfo {author} {\bibfnamefont {A.}~\bibnamefont {Albertini}},
  \bibinfo {author} {\bibfnamefont {M.}~\bibnamefont {Kuss}},\ and\ \bibinfo
  {author} {\bibfnamefont {R.}~\bibnamefont {Lindner}},\ }\bibfield  {title}
  {\bibinfo {title} {{Optical detection of symmetry breaking on a nanoscale in
  SrTiO$_3$:Ca}},\ }\href {https://doi.org/10.1080/00150199708012832}
  {\bibfield  {journal} {\bibinfo  {journal} {Ferroelectrics}\ }\textbf
  {\bibinfo {volume} {203}},\ \bibinfo {pages} {57} (\bibinfo {year}
  {1997})}\BibitemShut {NoStop}%
\bibitem [{\citenamefont {Carpenter}\ \emph {et~al.}(2006)\citenamefont
  {Carpenter}, \citenamefont {Howard}, \citenamefont {Knight},\ and\
  \citenamefont {Zhang}}]{Carpenter2006}%
  \BibitemOpen
  \bibfield  {author} {\bibinfo {author} {\bibfnamefont {M.~A.}\ \bibnamefont
  {Carpenter}}, \bibinfo {author} {\bibfnamefont {C.~J.}\ \bibnamefont
  {Howard}}, \bibinfo {author} {\bibfnamefont {K.~S.}\ \bibnamefont {Knight}},\
  and\ \bibinfo {author} {\bibfnamefont {Z.}~\bibnamefont {Zhang}},\ }\bibfield
   {title} {\bibinfo {title} {{Structural relationships and a phase diagram for
  (Ca,Sr)TiO$_3$ perovskites}},\ }\href
  {https://doi.org/10.1088/0953-8984/18/48/002} {\bibfield  {journal} {\bibinfo
   {journal} {J. Phys.: Condens. Matter}\ }\textbf {\bibinfo {volume} {18}},\
  \bibinfo {pages} {10725} (\bibinfo {year} {2006})}\BibitemShut {NoStop}%
\bibitem [{\citenamefont {Mishra}\ and\ \citenamefont
  {Pandey}(2009)}]{Mishra2009}%
  \BibitemOpen
  \bibfield  {author} {\bibinfo {author} {\bibfnamefont {S.~K.}\ \bibnamefont
  {Mishra}}\ and\ \bibinfo {author} {\bibfnamefont {D.}~\bibnamefont
  {Pandey}},\ }\bibfield  {title} {\bibinfo {title} {{Low temperature x-ray
  diffraction study of the phase transitions in Sr$_{1-x}$Ca$_x$TiO$_3$ ($x$ =
  0.02, 0.04): Evidence for ferrielectric ordering}},\ }\href
  {https://doi.org/10.1063/1.3273863} {\bibfield  {journal} {\bibinfo
  {journal} {Appl. Phys. Lett.}\ }\textbf {\bibinfo {volume} {95}},\ \bibinfo
  {pages} {232910} (\bibinfo {year} {2009})}\BibitemShut {NoStop}%
\bibitem [{\citenamefont {Frederikse}\ \emph {et~al.}(1964)\citenamefont
  {Frederikse}, \citenamefont {Thurber},\ and\ \citenamefont
  {Hosler}}]{Frederikse1964}%
  \BibitemOpen
  \bibfield  {author} {\bibinfo {author} {\bibfnamefont {H.~P.~R.}\
  \bibnamefont {Frederikse}}, \bibinfo {author} {\bibfnamefont {W.~R.}\
  \bibnamefont {Thurber}},\ and\ \bibinfo {author} {\bibfnamefont {W.~R.}\
  \bibnamefont {Hosler}},\ }\bibfield  {title} {\bibinfo {title} {{Electronic
  Transport in Strontium Titanate}},\ }\href
  {https://doi.org/10.1103/physrev.134.a442} {\bibfield  {journal} {\bibinfo
  {journal} {Phys. Rev.}\ }\textbf {\bibinfo {volume} {134}},\ \bibinfo {pages}
  {A442} (\bibinfo {year} {1964})}\BibitemShut {NoStop}%
\bibitem [{\citenamefont {Tufte}\ and\ \citenamefont
  {Chapman}(1967)}]{Tufte1967}%
  \BibitemOpen
  \bibfield  {author} {\bibinfo {author} {\bibfnamefont {O.~N.}\ \bibnamefont
  {Tufte}}\ and\ \bibinfo {author} {\bibfnamefont {P.~W.}\ \bibnamefont
  {Chapman}},\ }\bibfield  {title} {\bibinfo {title} {{Electron Mobility in
  Semiconducting Strontium Titanate}},\ }\href
  {https://doi.org/10.1103/physrev.155.796} {\bibfield  {journal} {\bibinfo
  {journal} {Phys. Rev.}\ }\textbf {\bibinfo {volume} {155}},\ \bibinfo {pages}
  {796} (\bibinfo {year} {1967})}\BibitemShut {NoStop}%
\bibitem [{\citenamefont {Frederikse}\ and\ \citenamefont
  {Hosler}(1967)}]{Frederikse1967}%
  \BibitemOpen
  \bibfield  {author} {\bibinfo {author} {\bibfnamefont {H.~P.~R.}\
  \bibnamefont {Frederikse}}\ and\ \bibinfo {author} {\bibfnamefont {W.~R.}\
  \bibnamefont {Hosler}},\ }\bibfield  {title} {\bibinfo {title} {{Hall
  mobility in SrTiO$_3$}},\ }\href {https://doi.org/10.1103/PhysRev.161.822}
  {\bibfield  {journal} {\bibinfo  {journal} {Phys. Rev.}\ }\textbf {\bibinfo
  {volume} {161}},\ \bibinfo {pages} {822} (\bibinfo {year}
  {1967})}\BibitemShut {NoStop}%
\bibitem [{\citenamefont {Lee}\ \emph {et~al.}(1975)\citenamefont {Lee},
  \citenamefont {Destry},\ and\ \citenamefont {Brebner}}]{Lee1975}%
  \BibitemOpen
  \bibfield  {author} {\bibinfo {author} {\bibfnamefont {C.}~\bibnamefont
  {Lee}}, \bibinfo {author} {\bibfnamefont {J.}~\bibnamefont {Destry}},\ and\
  \bibinfo {author} {\bibfnamefont {J.~L.}\ \bibnamefont {Brebner}},\
  }\bibfield  {title} {\bibinfo {title} {{Optical absorption and transport in
  semiconducting SrTiO$_3$}},\ }\href
  {https://doi.org/10.1103/PhysRevB.11.2299} {\bibfield  {journal} {\bibinfo
  {journal} {Phys. Rev. B}\ }\textbf {\bibinfo {volume} {11}},\ \bibinfo
  {pages} {2299} (\bibinfo {year} {1975})}\BibitemShut {NoStop}%
\bibitem [{\citenamefont {Spinelli}\ \emph {et~al.}(2010)\citenamefont
  {Spinelli}, \citenamefont {Torija}, \citenamefont {Liu}, \citenamefont
  {Jan},\ and\ \citenamefont {Leighton}}]{Spinelli2010}%
  \BibitemOpen
  \bibfield  {author} {\bibinfo {author} {\bibfnamefont {A.}~\bibnamefont
  {Spinelli}}, \bibinfo {author} {\bibfnamefont {M.~A.}\ \bibnamefont
  {Torija}}, \bibinfo {author} {\bibfnamefont {C.}~\bibnamefont {Liu}},
  \bibinfo {author} {\bibfnamefont {C.}~\bibnamefont {Jan}},\ and\ \bibinfo
  {author} {\bibfnamefont {C.}~\bibnamefont {Leighton}},\ }\bibfield  {title}
  {\bibinfo {title} {{Electronic transport in doped SrTiO$_3$: Conduction
  mechanisms and potential applications}},\ }\href
  {https://doi.org/10.1103/PhysRevB.81.155110} {\bibfield  {journal} {\bibinfo
  {journal} {Phys. Rev. B}\ }\textbf {\bibinfo {volume} {81}},\ \bibinfo
  {pages} {155110} (\bibinfo {year} {2010})}\BibitemShut {NoStop}%
\bibitem [{\citenamefont {Schooley}\ \emph {et~al.}(1964)\citenamefont
  {Schooley}, \citenamefont {Hosler},\ and\ \citenamefont
  {Cohen}}]{Schooley1964}%
  \BibitemOpen
  \bibfield  {author} {\bibinfo {author} {\bibfnamefont {J.~F.}\ \bibnamefont
  {Schooley}}, \bibinfo {author} {\bibfnamefont {W.~R.}\ \bibnamefont
  {Hosler}},\ and\ \bibinfo {author} {\bibfnamefont {M.~L.}\ \bibnamefont
  {Cohen}},\ }\bibfield  {title} {\bibinfo {title} {{Superconductivity in
  semiconducting SrTiO$_3$}},\ }\href
  {https://doi.org/10.1103/PhysRevLett.12.474} {\bibfield  {journal} {\bibinfo
  {journal} {Phys. Rev. Lett.}\ }\textbf {\bibinfo {volume} {12}},\ \bibinfo
  {pages} {474} (\bibinfo {year} {1964})}\BibitemShut {NoStop}%
\bibitem [{\citenamefont {Lin}\ \emph {et~al.}(2014)\citenamefont {Lin},
  \citenamefont {Bridoux}, \citenamefont {Gourgout}, \citenamefont {Seyfarth},
  \citenamefont {Kr{\"{a}}mer}, \citenamefont {Nardone}, \citenamefont
  {Fauqu{\'{e}}},\ and\ \citenamefont {Behnia}}]{Lin2014}%
  \BibitemOpen
  \bibfield  {author} {\bibinfo {author} {\bibfnamefont {X.}~\bibnamefont
  {Lin}}, \bibinfo {author} {\bibfnamefont {G.}~\bibnamefont {Bridoux}},
  \bibinfo {author} {\bibfnamefont {A.}~\bibnamefont {Gourgout}}, \bibinfo
  {author} {\bibfnamefont {G.}~\bibnamefont {Seyfarth}}, \bibinfo {author}
  {\bibfnamefont {S.}~\bibnamefont {Kr{\"{a}}mer}}, \bibinfo {author}
  {\bibfnamefont {M.}~\bibnamefont {Nardone}}, \bibinfo {author} {\bibfnamefont
  {B.}~\bibnamefont {Fauqu{\'{e}}}},\ and\ \bibinfo {author} {\bibfnamefont
  {K.}~\bibnamefont {Behnia}},\ }\bibfield  {title} {\bibinfo {title}
  {{Critical doping for the onset of a two-band superconducting ground state in
  SrTiO$_{3-\delta}$}},\ }\href
  {https://doi.org/10.1103/PhysRevLett.112.207002} {\bibfield  {journal}
  {\bibinfo  {journal} {Phys. Rev. Lett.}\ }\textbf {\bibinfo {volume} {112}},\
  \bibinfo {pages} {207002} (\bibinfo {year} {2014})}\BibitemShut {NoStop}%
\bibitem [{\citenamefont {{de Lima}}\ \emph {et~al.}(2015)\citenamefont {{de
  Lima}}, \citenamefont {{da Luz}}, \citenamefont {Oliveira}, \citenamefont
  {Alves}, \citenamefont {{dos Santos}}, \citenamefont {Jomard}, \citenamefont
  {Sidis}, \citenamefont {Bourges}, \citenamefont {Harms}, \citenamefont
  {Grams}, \citenamefont {Hemberger}, \citenamefont {Lin}, \citenamefont
  {Fauqu{\'{e}}},\ and\ \citenamefont {Behnia}}]{DeLima2015}%
  \BibitemOpen
  \bibfield  {author} {\bibinfo {author} {\bibfnamefont {B.~S.}\ \bibnamefont
  {{de Lima}}}, \bibinfo {author} {\bibfnamefont {M.~S.}\ \bibnamefont {{da
  Luz}}}, \bibinfo {author} {\bibfnamefont {F.~S.}\ \bibnamefont {Oliveira}},
  \bibinfo {author} {\bibfnamefont {L.~M.~S.}\ \bibnamefont {Alves}}, \bibinfo
  {author} {\bibfnamefont {C.~A.~M.}\ \bibnamefont {{dos Santos}}}, \bibinfo
  {author} {\bibfnamefont {F.}~\bibnamefont {Jomard}}, \bibinfo {author}
  {\bibfnamefont {Y.}~\bibnamefont {Sidis}}, \bibinfo {author} {\bibfnamefont
  {P.}~\bibnamefont {Bourges}}, \bibinfo {author} {\bibfnamefont
  {S.}~\bibnamefont {Harms}}, \bibinfo {author} {\bibfnamefont {C.~P.}\
  \bibnamefont {Grams}}, \bibinfo {author} {\bibfnamefont {J.}~\bibnamefont
  {Hemberger}}, \bibinfo {author} {\bibfnamefont {X.}~\bibnamefont {Lin}},
  \bibinfo {author} {\bibfnamefont {B.}~\bibnamefont {Fauqu{\'{e}}}},\ and\
  \bibinfo {author} {\bibfnamefont {K.}~\bibnamefont {Behnia}},\ }\bibfield
  {title} {\bibinfo {title} {{Interplay between antiferrodistortive,
  ferroelectric, and superconducting instabilities in
  Sr$_{1-x}$Ca$_x$TiO$_{3-\delta}$}},\ }\href
  {https://doi.org/10.1103/PhysRevB.91.045108} {\bibfield  {journal} {\bibinfo
  {journal} {Phys. Rev. B}\ }\textbf {\bibinfo {volume} {91}},\ \bibinfo
  {pages} {045108} (\bibinfo {year} {2015})}\BibitemShut {NoStop}%
\bibitem [{\citenamefont {Rischau}\ \emph {et~al.}(2017)\citenamefont
  {Rischau}, \citenamefont {Lin}, \citenamefont {Grams}, \citenamefont {Finck},
  \citenamefont {Harms}, \citenamefont {Engelmayer}, \citenamefont {Lorenz},
  \citenamefont {Gallais}, \citenamefont {Fauqu{\'{e}}}, \citenamefont
  {Hemberger},\ and\ \citenamefont {Behnia}}]{Rischau2017}%
  \BibitemOpen
  \bibfield  {author} {\bibinfo {author} {\bibfnamefont {C.~W.}\ \bibnamefont
  {Rischau}}, \bibinfo {author} {\bibfnamefont {X.}~\bibnamefont {Lin}},
  \bibinfo {author} {\bibfnamefont {C.~P.}\ \bibnamefont {Grams}}, \bibinfo
  {author} {\bibfnamefont {D.}~\bibnamefont {Finck}}, \bibinfo {author}
  {\bibfnamefont {S.}~\bibnamefont {Harms}}, \bibinfo {author} {\bibfnamefont
  {J.}~\bibnamefont {Engelmayer}}, \bibinfo {author} {\bibfnamefont
  {T.}~\bibnamefont {Lorenz}}, \bibinfo {author} {\bibfnamefont
  {Y.}~\bibnamefont {Gallais}}, \bibinfo {author} {\bibfnamefont
  {B.}~\bibnamefont {Fauqu{\'{e}}}}, \bibinfo {author} {\bibfnamefont
  {J.}~\bibnamefont {Hemberger}},\ and\ \bibinfo {author} {\bibfnamefont
  {K.}~\bibnamefont {Behnia}},\ }\bibfield  {title} {\bibinfo {title} {{A
  ferroelectric quantum phase transition inside the superconducting dome of
  Sr$_{1-x}$Ca$_x$TiO$_{3-\delta}$}},\ }\href
  {https://doi.org/10.1038/nphys4085} {\bibfield  {journal} {\bibinfo
  {journal} {Nat. Phys.}\ }\textbf {\bibinfo {volume} {13}},\ \bibinfo {pages}
  {643} (\bibinfo {year} {2017})}\BibitemShut {NoStop}%
\bibitem [{\citenamefont {Glinchuk}\ and\ \citenamefont
  {Kondakova}(1992)}]{Glinchuk1992}%
  \BibitemOpen
  \bibfield  {author} {\bibinfo {author} {\bibfnamefont {M.~D.}\ \bibnamefont
  {Glinchuk}}\ and\ \bibinfo {author} {\bibfnamefont {I.~V.}\ \bibnamefont
  {Kondakova}},\ }\bibfield  {title} {\bibinfo {title} {{Ruderman-Kittel-like
  interaction of electric dipoles in systems with carriers}},\ }\href
  {https://doi.org/10.1002/pssb.2221740119} {\bibfield  {journal} {\bibinfo
  {journal} {phys. stat. sol. (b)}\ }\textbf {\bibinfo {volume} {174}},\
  \bibinfo {pages} {193} (\bibinfo {year} {1992})}\BibitemShut {NoStop}%
\bibitem [{\citenamefont {Glinchuk}\ \emph {et~al.}(1994)\citenamefont
  {Glinchuk}, \citenamefont {Kondakova},\ and\ \citenamefont
  {Kuzian}}]{Glinchuk1994}%
  \BibitemOpen
  \bibfield  {author} {\bibinfo {author} {\bibfnamefont {M.~D.}\ \bibnamefont
  {Glinchuk}}, \bibinfo {author} {\bibfnamefont {I.~V.}\ \bibnamefont
  {Kondakova}},\ and\ \bibinfo {author} {\bibfnamefont {R.~O.}\ \bibnamefont
  {Kuzian}},\ }\bibfield  {title} {\bibinfo {title} {{The possibility of
  Kondo-like effect in systems with non-tunneling off-center ions}},\ }\href
  {https://doi.org/10.1080/00150199408016549} {\bibfield  {journal} {\bibinfo
  {journal} {Ferroelectrics}\ }\textbf {\bibinfo {volume} {153}},\ \bibinfo
  {pages} {97} (\bibinfo {year} {1994})}\BibitemShut {NoStop}%
\bibitem [{\citenamefont {Rowley}\ \emph {et~al.}(2014)\citenamefont {Rowley},
  \citenamefont {Spalek}, \citenamefont {Smith}, \citenamefont {Dean},
  \citenamefont {Itoh}, \citenamefont {Scott}, \citenamefont {Lonzarich},\ and\
  \citenamefont {Saxena}}]{Rowley2014}%
  \BibitemOpen
  \bibfield  {author} {\bibinfo {author} {\bibfnamefont {S.~E.}\ \bibnamefont
  {Rowley}}, \bibinfo {author} {\bibfnamefont {L.~J.}\ \bibnamefont {Spalek}},
  \bibinfo {author} {\bibfnamefont {R.~P.}\ \bibnamefont {Smith}}, \bibinfo
  {author} {\bibfnamefont {M.~P.~M.}\ \bibnamefont {Dean}}, \bibinfo {author}
  {\bibfnamefont {M.}~\bibnamefont {Itoh}}, \bibinfo {author} {\bibfnamefont
  {J.~F.}\ \bibnamefont {Scott}}, \bibinfo {author} {\bibfnamefont {G.~G.}\
  \bibnamefont {Lonzarich}},\ and\ \bibinfo {author} {\bibfnamefont {S.~S.}\
  \bibnamefont {Saxena}},\ }\bibfield  {title} {\bibinfo {title}
  {{Ferroelectric quantum criticality}},\ }\href
  {https://doi.org/10.1038/nphys2924} {\bibfield  {journal} {\bibinfo
  {journal} {Nat. Phys.}\ }\textbf {\bibinfo {volume} {10}},\ \bibinfo {pages}
  {367} (\bibinfo {year} {2014})}\BibitemShut {NoStop}%
\bibitem [{\citenamefont {Uwe}\ and\ \citenamefont {Sakudo}(1976)}]{Uwe1976}%
  \BibitemOpen
  \bibfield  {author} {\bibinfo {author} {\bibfnamefont {H.}~\bibnamefont
  {Uwe}}\ and\ \bibinfo {author} {\bibfnamefont {T.}~\bibnamefont {Sakudo}},\
  }\bibfield  {title} {\bibinfo {title} {{Stress-induced ferroelectricity and
  soft phonon modes in SrTiO$_3$}},\ }\href
  {https://doi.org/10.1103/physrevb.13.271} {\bibfield  {journal} {\bibinfo
  {journal} {Phys. Rev. B}\ }\textbf {\bibinfo {volume} {13}},\ \bibinfo
  {pages} {271} (\bibinfo {year} {1976})}\BibitemShut {NoStop}%
\bibitem [{\citenamefont {Itoh}\ \emph {et~al.}(1999)\citenamefont {Itoh},
  \citenamefont {Wang}, \citenamefont {Inaguma}, \citenamefont {Yamaguchi},
  \citenamefont {Shan},\ and\ \citenamefont {Nakamura}}]{Itoh1999}%
  \BibitemOpen
  \bibfield  {author} {\bibinfo {author} {\bibfnamefont {M.}~\bibnamefont
  {Itoh}}, \bibinfo {author} {\bibfnamefont {R.}~\bibnamefont {Wang}}, \bibinfo
  {author} {\bibfnamefont {Y.}~\bibnamefont {Inaguma}}, \bibinfo {author}
  {\bibfnamefont {T.}~\bibnamefont {Yamaguchi}}, \bibinfo {author}
  {\bibfnamefont {Y.-J.}\ \bibnamefont {Shan}},\ and\ \bibinfo {author}
  {\bibfnamefont {T.}~\bibnamefont {Nakamura}},\ }\bibfield  {title} {\bibinfo
  {title} {{Ferroelectricity Induced by Oxygen Isotope Exchange in Strontium
  Titanate Perovskite}},\ }\href {https://doi.org/10.1103/physrevlett.82.3540}
  {\bibfield  {journal} {\bibinfo  {journal} {Phys. Rev. Lett.}\ }\textbf
  {\bibinfo {volume} {82}},\ \bibinfo {pages} {3540} (\bibinfo {year}
  {1999})}\BibitemShut {NoStop}%
\bibitem [{\citenamefont {Zhu}\ \emph {et~al.}(2003)\citenamefont {Zhu},
  \citenamefont {Garst}, \citenamefont {Rosch},\ and\ \citenamefont
  {Si}}]{Zhu2003}%
  \BibitemOpen
  \bibfield  {author} {\bibinfo {author} {\bibfnamefont {L.}~\bibnamefont
  {Zhu}}, \bibinfo {author} {\bibfnamefont {M.}~\bibnamefont {Garst}}, \bibinfo
  {author} {\bibfnamefont {A.}~\bibnamefont {Rosch}},\ and\ \bibinfo {author}
  {\bibfnamefont {Q.}~\bibnamefont {Si}},\ }\bibfield  {title} {\bibinfo
  {title} {{Universally Diverging Gr{\"{u}}neisen Parameter and the
  Magnetocaloric Effect Close to Quantum Critical Points}},\ }\href
  {https://doi.org/10.1103/physrevlett.91.066404} {\bibfield  {journal}
  {\bibinfo  {journal} {Phys. Rev. Lett.}\ }\textbf {\bibinfo {volume} {91}},\
  \bibinfo {pages} {066404} (\bibinfo {year} {2003})}\BibitemShut {NoStop}%
\bibitem [{\citenamefont {Garst}\ and\ \citenamefont
  {Rosch}(2005)}]{Garst2005}%
  \BibitemOpen
  \bibfield  {author} {\bibinfo {author} {\bibfnamefont {M.}~\bibnamefont
  {Garst}}\ and\ \bibinfo {author} {\bibfnamefont {A.}~\bibnamefont {Rosch}},\
  }\bibfield  {title} {\bibinfo {title} {{Sign change of the Gr{\"{u}}neisen
  parameter and magnetocaloric effect near quantum critical points}},\ }\href
  {https://doi.org/10.1103/physrevb.72.205129} {\bibfield  {journal} {\bibinfo
  {journal} {Phys. Rev. B}\ }\textbf {\bibinfo {volume} {72}},\ \bibinfo
  {pages} {205129} (\bibinfo {year} {2005})}\BibitemShut {NoStop}%
\bibitem [{Note1()}]{Note1}%
  \BibitemOpen
  \bibinfo {note} {The divergence and sign change of $\alpha /c_p$ require a
  finite pressure dependence of the underlying quantum phase transition, which,
  for example, may result from a pressure-dependent quantum critical magnetic
  field.}\BibitemShut {Stop}%
\bibitem [{\citenamefont {K{\"{u}}chler}\ \emph {et~al.}(2003)\citenamefont
  {K{\"{u}}chler}, \citenamefont {Oeschler}, \citenamefont {Gegenwart},
  \citenamefont {Cichorek}, \citenamefont {Neumaier}, \citenamefont {Tegus},
  \citenamefont {Geibel}, \citenamefont {Mydosh}, \citenamefont {Steglich},
  \citenamefont {Zhu},\ and\ \citenamefont {Si}}]{Kuechler2003}%
  \BibitemOpen
  \bibfield  {author} {\bibinfo {author} {\bibfnamefont {R.}~\bibnamefont
  {K{\"{u}}chler}}, \bibinfo {author} {\bibfnamefont {N.}~\bibnamefont
  {Oeschler}}, \bibinfo {author} {\bibfnamefont {P.}~\bibnamefont {Gegenwart}},
  \bibinfo {author} {\bibfnamefont {T.}~\bibnamefont {Cichorek}}, \bibinfo
  {author} {\bibfnamefont {K.}~\bibnamefont {Neumaier}}, \bibinfo {author}
  {\bibfnamefont {O.}~\bibnamefont {Tegus}}, \bibinfo {author} {\bibfnamefont
  {C.}~\bibnamefont {Geibel}}, \bibinfo {author} {\bibfnamefont {J.~A.}\
  \bibnamefont {Mydosh}}, \bibinfo {author} {\bibfnamefont {F.}~\bibnamefont
  {Steglich}}, \bibinfo {author} {\bibfnamefont {L.}~\bibnamefont {Zhu}},\ and\
  \bibinfo {author} {\bibfnamefont {Q.}~\bibnamefont {Si}},\ }\bibfield
  {title} {\bibinfo {title} {{Divergence of the Gr{\"{u}}neisen Ratio at
  Quantum Critical Points in Heavy Fermion Metals}},\ }\href
  {https://doi.org/10.1103/physrevlett.91.066405} {\bibfield  {journal}
  {\bibinfo  {journal} {Phys. Rev. Lett.}\ }\textbf {\bibinfo {volume} {91}},\
  \bibinfo {pages} {066405} (\bibinfo {year} {2003})}\BibitemShut {NoStop}%
\bibitem [{\citenamefont {Gegenwart}\ \emph {et~al.}(2006)\citenamefont
  {Gegenwart}, \citenamefont {Weickert}, \citenamefont {Garst}, \citenamefont
  {Perry},\ and\ \citenamefont {Maeno}}]{Gegenwart2006}%
  \BibitemOpen
  \bibfield  {author} {\bibinfo {author} {\bibfnamefont {P.}~\bibnamefont
  {Gegenwart}}, \bibinfo {author} {\bibfnamefont {F.}~\bibnamefont {Weickert}},
  \bibinfo {author} {\bibfnamefont {M.}~\bibnamefont {Garst}}, \bibinfo
  {author} {\bibfnamefont {R.~S.}\ \bibnamefont {Perry}},\ and\ \bibinfo
  {author} {\bibfnamefont {Y.}~\bibnamefont {Maeno}},\ }\bibfield  {title}
  {\bibinfo {title} {{Metamagnetic Quantum Criticality in Sr$_3$Ru$_2$O$_7$
  Studied by Thermal Expansion}},\ }\href
  {https://doi.org/10.1103/physrevlett.96.136402} {\bibfield  {journal}
  {\bibinfo  {journal} {Phys. Rev. Lett.}\ }\textbf {\bibinfo {volume} {96}},\
  \bibinfo {pages} {136402} (\bibinfo {year} {2006})}\BibitemShut {NoStop}%
\bibitem [{\citenamefont {Baier}\ \emph {et~al.}(2007)\citenamefont {Baier},
  \citenamefont {Steffens}, \citenamefont {Schumann}, \citenamefont {Kriener},
  \citenamefont {Stark}, \citenamefont {Hartmann}, \citenamefont {Friedt},
  \citenamefont {Revcolevschi}, \citenamefont {Radaelli}, \citenamefont
  {Nakatsuji}, \citenamefont {Maeno}, \citenamefont {Mydosh}, \citenamefont
  {Lorenz},\ and\ \citenamefont {Braden}}]{Baier2007}%
  \BibitemOpen
  \bibfield  {author} {\bibinfo {author} {\bibfnamefont {J.}~\bibnamefont
  {Baier}}, \bibinfo {author} {\bibfnamefont {P.}~\bibnamefont {Steffens}},
  \bibinfo {author} {\bibfnamefont {O.}~\bibnamefont {Schumann}}, \bibinfo
  {author} {\bibfnamefont {M.}~\bibnamefont {Kriener}}, \bibinfo {author}
  {\bibfnamefont {S.}~\bibnamefont {Stark}}, \bibinfo {author} {\bibfnamefont
  {H.}~\bibnamefont {Hartmann}}, \bibinfo {author} {\bibfnamefont
  {O.}~\bibnamefont {Friedt}}, \bibinfo {author} {\bibfnamefont
  {A.}~\bibnamefont {Revcolevschi}}, \bibinfo {author} {\bibfnamefont {P.~G.}\
  \bibnamefont {Radaelli}}, \bibinfo {author} {\bibfnamefont {S.}~\bibnamefont
  {Nakatsuji}}, \bibinfo {author} {\bibfnamefont {Y.}~\bibnamefont {Maeno}},
  \bibinfo {author} {\bibfnamefont {J.~A.}\ \bibnamefont {Mydosh}}, \bibinfo
  {author} {\bibfnamefont {T.}~\bibnamefont {Lorenz}},\ and\ \bibinfo {author}
  {\bibfnamefont {M.}~\bibnamefont {Braden}},\ }\bibfield  {title} {\bibinfo
  {title} {{Magnetoelastic Coupling Across the Metamagnetic Transition in
  Ca$_{2-x}$Sr$_x$RuO$_4$ $(0.2\leq x\leq 0.5)$}},\ }\href
  {https://doi.org/10.1007/s10909-007-9330-0} {\bibfield  {journal} {\bibinfo
  {journal} {J. Low Temp. Phys.}\ }\textbf {\bibinfo {volume} {147}},\ \bibinfo
  {pages} {405} (\bibinfo {year} {2007})}\BibitemShut {NoStop}%
\bibitem [{\citenamefont {Lorenz}\ \emph {et~al.}(2007)\citenamefont {Lorenz},
  \citenamefont {Stark}, \citenamefont {Heyer}, \citenamefont {Hollmann},
  \citenamefont {Vasiliev}, \citenamefont {Oosawa},\ and\ \citenamefont
  {Tanaka}}]{Lorenz2007}%
  \BibitemOpen
  \bibfield  {author} {\bibinfo {author} {\bibfnamefont {T.}~\bibnamefont
  {Lorenz}}, \bibinfo {author} {\bibfnamefont {S.}~\bibnamefont {Stark}},
  \bibinfo {author} {\bibfnamefont {O.}~\bibnamefont {Heyer}}, \bibinfo
  {author} {\bibfnamefont {N.}~\bibnamefont {Hollmann}}, \bibinfo {author}
  {\bibfnamefont {A.}~\bibnamefont {Vasiliev}}, \bibinfo {author}
  {\bibfnamefont {A.}~\bibnamefont {Oosawa}},\ and\ \bibinfo {author}
  {\bibfnamefont {H.}~\bibnamefont {Tanaka}},\ }\bibfield  {title} {\bibinfo
  {title} {{Thermodynamics of the coupled spin-dimer system TlCuCl$_3$ close to
  a quantum phase transition}},\ }\href
  {https://doi.org/10.1016/j.jmmm.2007.02.154} {\bibfield  {journal} {\bibinfo
  {journal} {J. Magn. Magn. Mater.}\ }\textbf {\bibinfo {volume} {316}},\
  \bibinfo {pages} {291} (\bibinfo {year} {2007})}\BibitemShut {NoStop}%
\bibitem [{\citenamefont {Lorenz}\ \emph {et~al.}(2008)\citenamefont {Lorenz},
  \citenamefont {Heyer}, \citenamefont {Garst}, \citenamefont {Anfuso},
  \citenamefont {Rosch}, \citenamefont {R{\"{u}}egg},\ and\ \citenamefont
  {Kr{\"{a}}mer}}]{Lorenz2008}%
  \BibitemOpen
  \bibfield  {author} {\bibinfo {author} {\bibfnamefont {T.}~\bibnamefont
  {Lorenz}}, \bibinfo {author} {\bibfnamefont {O.}~\bibnamefont {Heyer}},
  \bibinfo {author} {\bibfnamefont {M.}~\bibnamefont {Garst}}, \bibinfo
  {author} {\bibfnamefont {F.}~\bibnamefont {Anfuso}}, \bibinfo {author}
  {\bibfnamefont {A.}~\bibnamefont {Rosch}}, \bibinfo {author} {\bibfnamefont
  {C.}~\bibnamefont {R{\"{u}}egg}},\ and\ \bibinfo {author} {\bibfnamefont
  {K.}~\bibnamefont {Kr{\"{a}}mer}},\ }\bibfield  {title} {\bibinfo {title}
  {{Diverging Thermal Expansion of the Spin-Ladder System
  (C$_5$H$_{12}$N)$_2$CuBr$_4$}},\ }\href
  {https://doi.org/10.1103/physrevlett.100.067208} {\bibfield  {journal}
  {\bibinfo  {journal} {Phys. Rev. Lett.}\ }\textbf {\bibinfo {volume} {100}},\
  \bibinfo {pages} {067208} (\bibinfo {year} {2008})}\BibitemShut {NoStop}%
\bibitem [{\citenamefont {Breunig}\ \emph {et~al.}(2013)\citenamefont
  {Breunig}, \citenamefont {Garst}, \citenamefont {Sela}, \citenamefont
  {Buldmann}, \citenamefont {Becker}, \citenamefont {Bohat{\'{y}}},
  \citenamefont {M{\"{u}}ller},\ and\ \citenamefont {Lorenz}}]{Breunig2013}%
  \BibitemOpen
  \bibfield  {author} {\bibinfo {author} {\bibfnamefont {O.}~\bibnamefont
  {Breunig}}, \bibinfo {author} {\bibfnamefont {M.}~\bibnamefont {Garst}},
  \bibinfo {author} {\bibfnamefont {E.}~\bibnamefont {Sela}}, \bibinfo {author}
  {\bibfnamefont {B.}~\bibnamefont {Buldmann}}, \bibinfo {author}
  {\bibfnamefont {P.}~\bibnamefont {Becker}}, \bibinfo {author} {\bibfnamefont
  {L.}~\bibnamefont {Bohat{\'{y}}}}, \bibinfo {author} {\bibfnamefont
  {R.}~\bibnamefont {M{\"{u}}ller}},\ and\ \bibinfo {author} {\bibfnamefont
  {T.}~\bibnamefont {Lorenz}},\ }\bibfield  {title} {\bibinfo {title}
  {{Spin-$\frac{1}{2}$ $XXZ$ Chain System Cs$_2$CoCl$_4$ in a Transverse
  Magnetic Field}},\ }\href {https://doi.org/10.1103/physrevlett.111.187202}
  {\bibfield  {journal} {\bibinfo  {journal} {Phys. Rev. Lett.}\ }\textbf
  {\bibinfo {volume} {111}},\ \bibinfo {pages} {187202} (\bibinfo {year}
  {2013})}\BibitemShut {NoStop}%
\bibitem [{\citenamefont {Breunig}\ \emph {et~al.}(2017)\citenamefont
  {Breunig}, \citenamefont {Garst}, \citenamefont {Kl{\"{u}}mper},
  \citenamefont {Rohrkamp}, \citenamefont {Turnbull},\ and\ \citenamefont
  {Lorenz}}]{Breunig2017a}%
  \BibitemOpen
  \bibfield  {author} {\bibinfo {author} {\bibfnamefont {O.}~\bibnamefont
  {Breunig}}, \bibinfo {author} {\bibfnamefont {M.}~\bibnamefont {Garst}},
  \bibinfo {author} {\bibfnamefont {A.}~\bibnamefont {Kl{\"{u}}mper}}, \bibinfo
  {author} {\bibfnamefont {J.}~\bibnamefont {Rohrkamp}}, \bibinfo {author}
  {\bibfnamefont {M.~M.}\ \bibnamefont {Turnbull}},\ and\ \bibinfo {author}
  {\bibfnamefont {T.}~\bibnamefont {Lorenz}},\ }\bibfield  {title} {\bibinfo
  {title} {{Quantum criticality in the spin-1/2 Heisenberg chain system copper
  pyrazine dinitrate}},\ }\href {https://doi.org/10.1126/sciadv.aao3773}
  {\bibfield  {journal} {\bibinfo  {journal} {Sci. Adv.}\ }\textbf {\bibinfo
  {volume} {3}},\ \bibinfo {pages} {eaao3773} (\bibinfo {year}
  {2017})}\BibitemShut {NoStop}%
\bibitem [{\citenamefont {Grube}\ \emph {et~al.}(2018)\citenamefont {Grube},
  \citenamefont {Pintschovius}, \citenamefont {Weber}, \citenamefont
  {Castellan}, \citenamefont {Zaum}, \citenamefont {Kuntz}, \citenamefont
  {Schweiss}, \citenamefont {Stockert}, \citenamefont {Bachus}, \citenamefont
  {Shimura}, \citenamefont {Fritsch},\ and\ \citenamefont
  {v.~L{\"{o}}hneysen}}]{Grube2018}%
  \BibitemOpen
  \bibfield  {author} {\bibinfo {author} {\bibfnamefont {K.}~\bibnamefont
  {Grube}}, \bibinfo {author} {\bibfnamefont {L.}~\bibnamefont {Pintschovius}},
  \bibinfo {author} {\bibfnamefont {F.}~\bibnamefont {Weber}}, \bibinfo
  {author} {\bibfnamefont {J.-P.}\ \bibnamefont {Castellan}}, \bibinfo {author}
  {\bibfnamefont {S.}~\bibnamefont {Zaum}}, \bibinfo {author} {\bibfnamefont
  {S.}~\bibnamefont {Kuntz}}, \bibinfo {author} {\bibfnamefont
  {P.}~\bibnamefont {Schweiss}}, \bibinfo {author} {\bibfnamefont
  {O.}~\bibnamefont {Stockert}}, \bibinfo {author} {\bibfnamefont
  {S.}~\bibnamefont {Bachus}}, \bibinfo {author} {\bibfnamefont
  {Y.}~\bibnamefont {Shimura}}, \bibinfo {author} {\bibfnamefont
  {V.}~\bibnamefont {Fritsch}},\ and\ \bibinfo {author} {\bibfnamefont
  {H.}~\bibnamefont {v.~L{\"{o}}hneysen}},\ }\bibfield  {title} {\bibinfo
  {title} {{Magnetic and Structural Quantum Phase Transitions in
  CeCu$_{6-x}$Au$_x$ are Independent}},\ }\href
  {https://doi.org/10.1103/physrevlett.121.087203} {\bibfield  {journal}
  {\bibinfo  {journal} {Phys. Rev. Lett.}\ }\textbf {\bibinfo {volume} {121}},\
  \bibinfo {pages} {087203} (\bibinfo {year} {2018})}\BibitemShut {NoStop}%
\bibitem [{\citenamefont {Meingast}\ \emph {et~al.}(2012)\citenamefont
  {Meingast}, \citenamefont {Hardy}, \citenamefont {Heid}, \citenamefont
  {Adelmann}, \citenamefont {B{\"{o}}hmer}, \citenamefont {Burger},
  \citenamefont {Ernst}, \citenamefont {Fromknecht}, \citenamefont {Schweiss},\
  and\ \citenamefont {Wolf}}]{Meingast2012}%
  \BibitemOpen
  \bibfield  {author} {\bibinfo {author} {\bibfnamefont {C.}~\bibnamefont
  {Meingast}}, \bibinfo {author} {\bibfnamefont {F.}~\bibnamefont {Hardy}},
  \bibinfo {author} {\bibfnamefont {R.}~\bibnamefont {Heid}}, \bibinfo {author}
  {\bibfnamefont {P.}~\bibnamefont {Adelmann}}, \bibinfo {author}
  {\bibfnamefont {A.}~\bibnamefont {B{\"{o}}hmer}}, \bibinfo {author}
  {\bibfnamefont {P.}~\bibnamefont {Burger}}, \bibinfo {author} {\bibfnamefont
  {D.}~\bibnamefont {Ernst}}, \bibinfo {author} {\bibfnamefont
  {R.}~\bibnamefont {Fromknecht}}, \bibinfo {author} {\bibfnamefont
  {P.}~\bibnamefont {Schweiss}},\ and\ \bibinfo {author} {\bibfnamefont
  {T.}~\bibnamefont {Wolf}},\ }\bibfield  {title} {\bibinfo {title} {{Thermal
  Expansion and Gr{\"{u}}neisen Parameters of Ba(Fe$_{1-x}$Co$_x$)$_2$As$_2$: A
  Thermodynamic Quest for Quantum Criticality}},\ }\href
  {https://doi.org/10.1103/physrevlett.108.177004} {\bibfield  {journal}
  {\bibinfo  {journal} {Phys. Rev. Lett.}\ }\textbf {\bibinfo {volume} {108}},\
  \bibinfo {pages} {177004} (\bibinfo {year} {2012})}\BibitemShut {NoStop}%
\bibitem [{\citenamefont {McCalla}\ \emph {et~al.}(2016)\citenamefont
  {McCalla}, \citenamefont {Walter},\ and\ \citenamefont
  {Leighton}}]{McCalla2016}%
  \BibitemOpen
  \bibfield  {author} {\bibinfo {author} {\bibfnamefont {E.}~\bibnamefont
  {McCalla}}, \bibinfo {author} {\bibfnamefont {J.}~\bibnamefont {Walter}},\
  and\ \bibinfo {author} {\bibfnamefont {C.}~\bibnamefont {Leighton}},\
  }\bibfield  {title} {\bibinfo {title} {{A Unified View of the
  Substitution-Dependent Antiferrodistortive Phase Transition in SrTiO$_3$}},\
  }\href {https://doi.org/10.1021/acs.chemmater.6b03667} {\bibfield  {journal}
  {\bibinfo  {journal} {Chem. Mater.}\ }\textbf {\bibinfo {volume} {28}},\
  \bibinfo {pages} {7973} (\bibinfo {year} {2016})}\BibitemShut {NoStop}%
\bibitem [{\citenamefont {Mishra}\ \emph {et~al.}(2005)\citenamefont {Mishra},
  \citenamefont {Ranjan}, \citenamefont {Pandey}, \citenamefont {Ranson},
  \citenamefont {Ouillon}, \citenamefont {Pinan-Lucarre},\ and\ \citenamefont
  {Pruzan}}]{Mishra2005}%
  \BibitemOpen
  \bibfield  {author} {\bibinfo {author} {\bibfnamefont {S.~K.}\ \bibnamefont
  {Mishra}}, \bibinfo {author} {\bibfnamefont {R.}~\bibnamefont {Ranjan}},
  \bibinfo {author} {\bibfnamefont {D.}~\bibnamefont {Pandey}}, \bibinfo
  {author} {\bibfnamefont {P.}~\bibnamefont {Ranson}}, \bibinfo {author}
  {\bibfnamefont {R.}~\bibnamefont {Ouillon}}, \bibinfo {author} {\bibfnamefont
  {J.-P.}\ \bibnamefont {Pinan-Lucarre}},\ and\ \bibinfo {author}
  {\bibfnamefont {P.}~\bibnamefont {Pruzan}},\ }\bibfield  {title} {\bibinfo
  {title} {{A combined X-ray diffraction and Raman scattering study of the
  phase transitions in Sr$_{1-x}$Ca$_x$TiO$_3$ ($x$ = 0.04, 0.06, and 0.12)}},\
  }\href {https://doi.org/10.1016/j.jssc.2005.06.036} {\bibfield  {journal}
  {\bibinfo  {journal} {J. Solid State Chem.}\ }\textbf {\bibinfo {volume}
  {178}},\ \bibinfo {pages} {2846} (\bibinfo {year} {2005})}\BibitemShut
  {NoStop}%
\bibitem [{\citenamefont {Rimai}\ and\ \citenamefont
  {deMars}(1962)}]{Rimai1962}%
  \BibitemOpen
  \bibfield  {author} {\bibinfo {author} {\bibfnamefont {L.}~\bibnamefont
  {Rimai}}\ and\ \bibinfo {author} {\bibfnamefont {G.~A.}\ \bibnamefont
  {deMars}},\ }\bibfield  {title} {\bibinfo {title} {{Electron Paramagnetic
  Resonance of Trivalent Gadolinium Ions in Strontium and Barium Titanates}},\
  }\href {https://doi.org/10.1103/PhysRev.127.702} {\bibfield  {journal}
  {\bibinfo  {journal} {Phys. Rev.}\ }\textbf {\bibinfo {volume} {127}},\
  \bibinfo {pages} {702} (\bibinfo {year} {1962})}\BibitemShut {NoStop}%
\bibitem [{\citenamefont {Lytle}(1964)}]{Lytle1964}%
  \BibitemOpen
  \bibfield  {author} {\bibinfo {author} {\bibfnamefont {F.~W.}\ \bibnamefont
  {Lytle}},\ }\bibfield  {title} {\bibinfo {title} {{X-Ray Diffractometry of
  Low-Temperature Phase Transformations in Strontium Titanate}},\ }\href
  {https://doi.org/10.1063/1.1702820} {\bibfield  {journal} {\bibinfo
  {journal} {J. Appl. Phys.}\ }\textbf {\bibinfo {volume} {35}},\ \bibinfo
  {pages} {2212} (\bibinfo {year} {1964})}\BibitemShut {NoStop}%
\bibitem [{\citenamefont {Ohama}\ \emph {et~al.}(1984)\citenamefont {Ohama},
  \citenamefont {Sakashita},\ and\ \citenamefont {Okazaki}}]{Ohama1984}%
  \BibitemOpen
  \bibfield  {author} {\bibinfo {author} {\bibfnamefont {N.}~\bibnamefont
  {Ohama}}, \bibinfo {author} {\bibfnamefont {H.}~\bibnamefont {Sakashita}},\
  and\ \bibinfo {author} {\bibfnamefont {A.}~\bibnamefont {Okazaki}},\
  }\bibfield  {title} {\bibinfo {title} {{The temperature dependence of the
  lattice constant of SrTiO$_3$ around the 105 K transition}},\ }\href
  {https://doi.org/10.1080/01411598408220325} {\bibfield  {journal} {\bibinfo
  {journal} {Phase Transit.}\ }\textbf {\bibinfo {volume} {4}},\ \bibinfo
  {pages} {81} (\bibinfo {year} {1984})}\BibitemShut {NoStop}%
\bibitem [{\citenamefont {Sato}\ \emph {et~al.}(1985)\citenamefont {Sato},
  \citenamefont {Soejima}, \citenamefont {Ohama}, \citenamefont {Okazaki},
  \citenamefont {Scheel},\ and\ \citenamefont {M{\"{u}}ller}}]{Sato1985}%
  \BibitemOpen
  \bibfield  {author} {\bibinfo {author} {\bibfnamefont {M.}~\bibnamefont
  {Sato}}, \bibinfo {author} {\bibfnamefont {Y.}~\bibnamefont {Soejima}},
  \bibinfo {author} {\bibfnamefont {N.}~\bibnamefont {Ohama}}, \bibinfo
  {author} {\bibfnamefont {A.}~\bibnamefont {Okazaki}}, \bibinfo {author}
  {\bibfnamefont {H.~J.}\ \bibnamefont {Scheel}},\ and\ \bibinfo {author}
  {\bibfnamefont {K.~A.}\ \bibnamefont {M{\"{u}}ller}},\ }\bibfield  {title}
  {\bibinfo {title} {{The lattice constant vs. temperature relation around the
  105 K transition of a flux-grown SrTiO$_3$ crystal}},\ }\href
  {https://doi.org/10.1080/01411598508209319} {\bibfield  {journal} {\bibinfo
  {journal} {Phase Transit.}\ }\textbf {\bibinfo {volume} {5}},\ \bibinfo
  {pages} {207} (\bibinfo {year} {1985})}\BibitemShut {NoStop}%
\bibitem [{\citenamefont {Mitsui}\ and\ \citenamefont
  {Westphal}(1961)}]{Mitsui1961}%
  \BibitemOpen
  \bibfield  {author} {\bibinfo {author} {\bibfnamefont {T.}~\bibnamefont
  {Mitsui}}\ and\ \bibinfo {author} {\bibfnamefont {W.~B.}\ \bibnamefont
  {Westphal}},\ }\bibfield  {title} {\bibinfo {title} {{Dielectric and X-Ray
  Studies of Ca$_x$Ba$_{1-x}$TiO$_3$ and Ca$_x$Sr$_{1-x}$TiO$_3$}},\ }\href
  {https://doi.org/10.1103/physrev.124.1354} {\bibfield  {journal} {\bibinfo
  {journal} {Phys. Rev.}\ }\textbf {\bibinfo {volume} {124}},\ \bibinfo {pages}
  {1354} (\bibinfo {year} {1961})}\BibitemShut {NoStop}%
\bibitem [{\citenamefont {Lemanov}(1997)}]{Lemanov1997}%
  \BibitemOpen
  \bibfield  {author} {\bibinfo {author} {\bibfnamefont {V.~V.}\ \bibnamefont
  {Lemanov}},\ }\bibfield  {title} {\bibinfo {title} {{Phase transitions in
  SrTiO$_3$-based solid solutions}},\ }\href
  {https://doi.org/10.1134/1.1130100} {\bibfield  {journal} {\bibinfo
  {journal} {Phys. Solid State}\ }\textbf {\bibinfo {volume} {39}},\ \bibinfo
  {pages} {1468} (\bibinfo {year} {1997})}\BibitemShut {NoStop}%
\bibitem [{\citenamefont {Unoki}\ and\ \citenamefont
  {Sakudo}(1967)}]{Unoki1967}%
  \BibitemOpen
  \bibfield  {author} {\bibinfo {author} {\bibfnamefont {H.}~\bibnamefont
  {Unoki}}\ and\ \bibinfo {author} {\bibfnamefont {T.}~\bibnamefont {Sakudo}},\
  }\bibfield  {title} {\bibinfo {title} {{Electron Spin Resonance of Fe$^{3+}$
  in SrTiO$_3$ with Special Reference to the 110$^{\circ}$K Phase
  Transition}},\ }\href {https://doi.org/10.1143/JPSJ.23.546} {\bibfield
  {journal} {\bibinfo  {journal} {J. Phys. Soc. Jpn.}\ }\textbf {\bibinfo
  {volume} {23}},\ \bibinfo {pages} {546} (\bibinfo {year} {1967})}\BibitemShut
  {NoStop}%
\bibitem [{\citenamefont {Fleury}\ \emph {et~al.}(1968)\citenamefont {Fleury},
  \citenamefont {Scott},\ and\ \citenamefont {Worlock}}]{Fleury1968}%
  \BibitemOpen
  \bibfield  {author} {\bibinfo {author} {\bibfnamefont {P.~A.}\ \bibnamefont
  {Fleury}}, \bibinfo {author} {\bibfnamefont {J.~F.}\ \bibnamefont {Scott}},\
  and\ \bibinfo {author} {\bibfnamefont {J.~M.}\ \bibnamefont {Worlock}},\
  }\bibfield  {title} {\bibinfo {title} {{Soft Phonon Modes and the
  110$^{\circ}$K Phase Transition in SrTiO$_3$}},\ }\href
  {https://doi.org/10.1103/physrevlett.21.16} {\bibfield  {journal} {\bibinfo
  {journal} {Phys. Rev. Lett.}\ }\textbf {\bibinfo {volume} {21}},\ \bibinfo
  {pages} {16} (\bibinfo {year} {1968})}\BibitemShut {NoStop}%
\bibitem [{\citenamefont {Shirane}\ and\ \citenamefont
  {Yamada}(1969)}]{Shirane1969}%
  \BibitemOpen
  \bibfield  {author} {\bibinfo {author} {\bibfnamefont {G.}~\bibnamefont
  {Shirane}}\ and\ \bibinfo {author} {\bibfnamefont {Y.}~\bibnamefont
  {Yamada}},\ }\bibfield  {title} {\bibinfo {title} {{Lattice-Dynamical Study
  of the 110$^{\circ}$K Phase Transition in SrTiO$_3$}},\ }\href
  {https://doi.org/10.1103/physrev.177.858} {\bibfield  {journal} {\bibinfo
  {journal} {Phys. Rev.}\ }\textbf {\bibinfo {volume} {177}},\ \bibinfo {pages}
  {858} (\bibinfo {year} {1969})}\BibitemShut {NoStop}%
\bibitem [{\citenamefont {Glazer}(1972)}]{Glazer1972}%
  \BibitemOpen
  \bibfield  {author} {\bibinfo {author} {\bibfnamefont {A.~M.}\ \bibnamefont
  {Glazer}},\ }\bibfield  {title} {\bibinfo {title} {{The classification of
  tilted octahedra in perovskites}},\ }\href
  {https://doi.org/10.1107/S0567740872007976} {\bibfield  {journal} {\bibinfo
  {journal} {Acta Crystallogr. Sect. B}\ }\textbf {\bibinfo {volume} {28}},\
  \bibinfo {pages} {3384} (\bibinfo {year} {1972})}\BibitemShut {NoStop}%
\bibitem [{\citenamefont {Glazer}(1975)}]{Glazer1975}%
  \BibitemOpen
  \bibfield  {author} {\bibinfo {author} {\bibfnamefont {A.~M.}\ \bibnamefont
  {Glazer}},\ }\bibfield  {title} {\bibinfo {title} {{Simple ways of
  determining perovskite structures}},\ }\href
  {https://doi.org/10.1107/s0567739475001635} {\bibfield  {journal} {\bibinfo
  {journal} {Acta Crystallogr. Sect. A}\ }\textbf {\bibinfo {volume} {31}},\
  \bibinfo {pages} {756} (\bibinfo {year} {1975})}\BibitemShut {NoStop}%
\bibitem [{\citenamefont {Cahn}(1954)}]{Cahn1954}%
  \BibitemOpen
  \bibfield  {author} {\bibinfo {author} {\bibfnamefont {R.~W.}\ \bibnamefont
  {Cahn}},\ }\bibfield  {title} {\bibinfo {title} {{Twinned crystals}},\ }\href
  {https://doi.org/10.1080/00018735400101223} {\bibfield  {journal} {\bibinfo
  {journal} {Adv. Phys.}\ }\textbf {\bibinfo {volume} {3}},\ \bibinfo {pages}
  {363} (\bibinfo {year} {1954})}\BibitemShut {NoStop}%
\bibitem [{\citenamefont {Authier}(2003)}]{Authier2003}%
  \BibitemOpen
  \bibinfo {editor} {\bibfnamefont {A.}~\bibnamefont {Authier}},\ ed.,\ \href
  {https://doi.org/10.1107/97809553602060000104} {\emph {\bibinfo {title}
  {{International Tables for Crystallography, Volume D}}}},\ \bibinfo {edition}
  {1st}\ ed.\ (\bibinfo  {publisher} {International Union of Crystallography},\
  \bibinfo {year} {2003})\BibitemShut {NoStop}%
\bibitem [{\citenamefont {Niesen}\ \emph {et~al.}(2013)\citenamefont {Niesen},
  \citenamefont {Kolland}, \citenamefont {Seher}, \citenamefont {Breunig},
  \citenamefont {Valldor}, \citenamefont {Braden}, \citenamefont {Grenier},\
  and\ \citenamefont {Lorenz}}]{Niesen2013}%
  \BibitemOpen
  \bibfield  {author} {\bibinfo {author} {\bibfnamefont {S.~K.}\ \bibnamefont
  {Niesen}}, \bibinfo {author} {\bibfnamefont {G.}~\bibnamefont {Kolland}},
  \bibinfo {author} {\bibfnamefont {M.}~\bibnamefont {Seher}}, \bibinfo
  {author} {\bibfnamefont {O.}~\bibnamefont {Breunig}}, \bibinfo {author}
  {\bibfnamefont {M.}~\bibnamefont {Valldor}}, \bibinfo {author} {\bibfnamefont
  {M.}~\bibnamefont {Braden}}, \bibinfo {author} {\bibfnamefont
  {B.}~\bibnamefont {Grenier}},\ and\ \bibinfo {author} {\bibfnamefont
  {T.}~\bibnamefont {Lorenz}},\ }\bibfield  {title} {\bibinfo {title}
  {{Magnetic phase diagrams, domain switching, and quantum phase transition of
  the quasi-one-dimensional Ising-like antiferromagnet BaCo$_2$V$_2$O$_8$}},\
  }\href {https://doi.org/10.1103/physrevb.87.224413} {\bibfield  {journal}
  {\bibinfo  {journal} {Phys. Rev. B}\ }\textbf {\bibinfo {volume} {87}},\
  \bibinfo {pages} {224413} (\bibinfo {year} {2013})}\BibitemShut {NoStop}%
\bibitem [{\citenamefont {Kunkem{\"{o}}ller}\ \emph {et~al.}(2017)\citenamefont
  {Kunkem{\"{o}}ller}, \citenamefont {Br{\"{u}}ning}, \citenamefont {Stunault},
  \citenamefont {Nugroho}, \citenamefont {Lorenz},\ and\ \citenamefont
  {Braden}}]{Kunkemoeller2017}%
  \BibitemOpen
  \bibfield  {author} {\bibinfo {author} {\bibfnamefont {S.}~\bibnamefont
  {Kunkem{\"{o}}ller}}, \bibinfo {author} {\bibfnamefont {D.}~\bibnamefont
  {Br{\"{u}}ning}}, \bibinfo {author} {\bibfnamefont {A.}~\bibnamefont
  {Stunault}}, \bibinfo {author} {\bibfnamefont {A.~A.}\ \bibnamefont
  {Nugroho}}, \bibinfo {author} {\bibfnamefont {T.}~\bibnamefont {Lorenz}},\
  and\ \bibinfo {author} {\bibfnamefont {M.}~\bibnamefont {Braden}},\
  }\bibfield  {title} {\bibinfo {title} {{Magnetic shape-memory effect in
  SrRuO$_3$}},\ }\href {https://doi.org/10.1103/physrevb.96.220406} {\bibfield
  {journal} {\bibinfo  {journal} {Phys. Rev. B}\ }\textbf {\bibinfo {volume}
  {96}},\ \bibinfo {pages} {220406(R)} (\bibinfo {year} {2017})}\BibitemShut
  {NoStop}%
\bibitem [{Note2()}]{Note2}%
  \BibitemOpen
  \bibinfo {note} {When measuring along $L_\protect \mathrm {long}$, the force
  applied via the dilatometer acts on the smallest cross-sectional area of our
  sample and consequently produces the largest pressure, whereas along
  $L_\protect \mathrm {short}$ the force acts on the largest cross section
  producing the smallest pressure. Nevertheless, we observe an expansion along
  $L_\protect \mathrm {long}$ and a compression along $L_\protect \mathrm
  {short}$ when cooling across $T_s$, which indicates the absence of
  stress-induced detwinning.}\BibitemShut {Stop}%
\bibitem [{\citenamefont {Kleemann}\ \emph {et~al.}(1988)\citenamefont
  {Kleemann}, \citenamefont {Sch{\"{a}}fer}, \citenamefont {M{\"{u}}ller},\
  and\ \citenamefont {Bednorz}}]{Kleemann1988}%
  \BibitemOpen
  \bibfield  {author} {\bibinfo {author} {\bibfnamefont {W.}~\bibnamefont
  {Kleemann}}, \bibinfo {author} {\bibfnamefont {F.~J.}\ \bibnamefont
  {Sch{\"{a}}fer}}, \bibinfo {author} {\bibfnamefont {K.~A.}\ \bibnamefont
  {M{\"{u}}ller}},\ and\ \bibinfo {author} {\bibfnamefont {J.~G.}\ \bibnamefont
  {Bednorz}},\ }\bibfield  {title} {\bibinfo {title} {{Domain state properties
  of the random-field xy-model system Sr$_{1-x}$Ca$_x$TiO$_3$}},\ }\href
  {https://doi.org/10.1080/00150198808223317} {\bibfield  {journal} {\bibinfo
  {journal} {Ferroelectrics}\ }\textbf {\bibinfo {volume} {80}},\ \bibinfo
  {pages} {297} (\bibinfo {year} {1988})}\BibitemShut {NoStop}%
\bibitem [{\citenamefont {B{\"{a}}uerle}\ and\ \citenamefont
  {Rehwald}(1978)}]{Baeuerle1978a}%
  \BibitemOpen
  \bibfield  {author} {\bibinfo {author} {\bibfnamefont {D.}~\bibnamefont
  {B{\"{a}}uerle}}\ and\ \bibinfo {author} {\bibfnamefont {W.}~\bibnamefont
  {Rehwald}},\ }\bibfield  {title} {\bibinfo {title} {{Structural phase
  transitions in semiconducting SrTiO$_3$}},\ }\href
  {https://doi.org/10.1016/0038-1098(78)91568-5} {\bibfield  {journal}
  {\bibinfo  {journal} {Solid State Commun.}\ }\textbf {\bibinfo {volume}
  {27}},\ \bibinfo {pages} {1343} (\bibinfo {year} {1978})}\BibitemShut
  {NoStop}%
\bibitem [{\citenamefont {Wagner}\ \emph {et~al.}(1980)\citenamefont {Wagner},
  \citenamefont {B{\"{a}}uerle}, \citenamefont {Schwabl}, \citenamefont
  {Dorner},\ and\ \citenamefont {Kraxenberger}}]{Wagner1980}%
  \BibitemOpen
  \bibfield  {author} {\bibinfo {author} {\bibfnamefont {D.}~\bibnamefont
  {Wagner}}, \bibinfo {author} {\bibfnamefont {D.}~\bibnamefont
  {B{\"{a}}uerle}}, \bibinfo {author} {\bibfnamefont {F.}~\bibnamefont
  {Schwabl}}, \bibinfo {author} {\bibfnamefont {B.}~\bibnamefont {Dorner}},\
  and\ \bibinfo {author} {\bibfnamefont {H.}~\bibnamefont {Kraxenberger}},\
  }\bibfield  {title} {\bibinfo {title} {{Soft modes in semiconducting
  SrTiO$_3$ -- I. The Zone Boundary Mode}},\ }\href
  {https://doi.org/10.1007/bf01352741} {\bibfield  {journal} {\bibinfo
  {journal} {Z. Phys. B}\ }\textbf {\bibinfo {volume} {37}},\ \bibinfo {pages}
  {317} (\bibinfo {year} {1980})}\BibitemShut {NoStop}%
\bibitem [{\citenamefont {Tao}\ \emph {et~al.}(2016)\citenamefont {Tao},
  \citenamefont {Loret}, \citenamefont {Xu}, \citenamefont {Yang},
  \citenamefont {Rischau}, \citenamefont {Lin}, \citenamefont {Fauqu{\'{e}}},
  \citenamefont {Verstraete},\ and\ \citenamefont {Behnia}}]{Tao2016}%
  \BibitemOpen
  \bibfield  {author} {\bibinfo {author} {\bibfnamefont {Q.}~\bibnamefont
  {Tao}}, \bibinfo {author} {\bibfnamefont {B.}~\bibnamefont {Loret}}, \bibinfo
  {author} {\bibfnamefont {B.}~\bibnamefont {Xu}}, \bibinfo {author}
  {\bibfnamefont {X.}~\bibnamefont {Yang}}, \bibinfo {author} {\bibfnamefont
  {C.~W.}\ \bibnamefont {Rischau}}, \bibinfo {author} {\bibfnamefont
  {X.}~\bibnamefont {Lin}}, \bibinfo {author} {\bibfnamefont {B.}~\bibnamefont
  {Fauqu{\'{e}}}}, \bibinfo {author} {\bibfnamefont {M.~J.}\ \bibnamefont
  {Verstraete}},\ and\ \bibinfo {author} {\bibfnamefont {K.}~\bibnamefont
  {Behnia}},\ }\bibfield  {title} {\bibinfo {title} {{Nonmonotonic anisotropy
  in charge conduction induced by antiferrodistortive transition in metallic
  SrTiO$_3$}},\ }\href {https://doi.org/10.1103/physrevb.94.035111} {\bibfield
  {journal} {\bibinfo  {journal} {Phys. Rev. B}\ }\textbf {\bibinfo {volume}
  {94}},\ \bibinfo {pages} {035111} (\bibinfo {year} {2016})}\BibitemShut
  {NoStop}%
\bibitem [{Note3()}]{Note3}%
  \BibitemOpen
  \bibinfo {note} {The decrease of $T_s$ is, however, not a direct consequence
  of the charge-carrier doping, because $n$-type doping by chemical
  substitution like in SrTi$_{1-x}$Nb$_x$O$_3$ increases $T_s$~\cite
  {Tao2016,McCalla2016}}\BibitemShut {NoStop}%
\bibitem [{Note4()}]{Note4}%
  \BibitemOpen
  \bibinfo {note} {On our twinned SrTiO$_3$ crystal we find a smooth $\alpha
  (T)$ low-temperature behavior. In contrast, thermal-expansion data on
  SrTiO$_3$ measured under strong uniaxial compressive stress result in
  single-domain tetragonal samples with very different expansion~\cite
  {Tsunekawa1984}. The corresponding $\alpha (T)$ curves show a minimum between
  \SI {20}{K} and \SI {30}{K}, which resembles the data of our Ca-doped
  samples. Because SrTiO$_3$ is known to develop a stress-induced ferroelectric
  order~\cite {Uwe1976}, the anomalous $\alpha (T)$ curves observed in~\cite
  {Tsunekawa1984} probably arise from ferroelectric order that is induced in
  those SrTiO$_3$ crystals by the application of the uniaxial compressive
  stress.}\BibitemShut {Stop}%
\bibitem [{Note5()}]{Note5}%
  \BibitemOpen
  \bibinfo {note} {This follows from general thermodynamics via
  Clausius-Clapeyron (Ehrenfest) relations for first- (second-)order phase
  transitions}\BibitemShut {NoStop}%
\bibitem [{\citenamefont {Tsunekawa}\ \emph {et~al.}(1984)\citenamefont
  {Tsunekawa}, \citenamefont {Watanabe},\ and\ \citenamefont
  {Takei}}]{Tsunekawa1984}%
  \BibitemOpen
  \bibfield  {author} {\bibinfo {author} {\bibfnamefont {S.}~\bibnamefont
  {Tsunekawa}}, \bibinfo {author} {\bibfnamefont {H.~F.~J.}\ \bibnamefont
  {Watanabe}},\ and\ \bibinfo {author} {\bibfnamefont {H.}~\bibnamefont
  {Takei}},\ }\bibfield  {title} {\bibinfo {title} {{Linear thermal expansion
  of SrTiO$_3$}},\ }\href {https://doi.org/10.1002/pssa.2210830207} {\bibfield
  {journal} {\bibinfo  {journal} {phys. stat. sol. (a)}\ }\textbf {\bibinfo
  {volume} {83}},\ \bibinfo {pages} {467} (\bibinfo {year} {1984})}\BibitemShut
  {NoStop}%
\end{thebibliography}
\end{document}